\documentclass[10pt]{amsart}

\usepackage{color}
\usepackage{amsmath}
\usepackage{amssymb}
\usepackage{float}
\usepackage{graphicx}

\newtheorem{proposition}{Proposition}[section]

\newtheorem{note}{Remark}[section]

\newsavebox{\savepar}

\def\G{\mathcal{G}}

\def\epsilon{\varepsilon}
\def\leq{\leqslant}
\def\geq{\geqslant}

\def\k{{\bf k}}

\def\ms{{\medskip\noindent}}
\def\bs{{\bigskip\noindent}}

\def\int{{\rm int}}

\newcommand{\rem}[1]{}

\begin{document}
\title{Agent Based Models and Opinion Dynamics as Markov Chains}
\author{Sven Banisch$^{a,d}$, Ricardo Lima$^{b,d}$ and Tanya Ara\'{u}jo$^{c,d}$}

\maketitle

\begin{center}
\begin{minipage}{13truecm}
\begin{small}

($^{A}$) Mathematical Physics, Bielefeld University (Germany)\footnote{Corresponding author:
Sven Banisch,
Mathematical Physics,
Physics Faculty,
Bielefeld University,
Postfach 100131,
D-33501 Bielefeld,
Germany. 
eMail: sven.banisch@UniVerseCity.de,
phone: 0049-0-521-106 6192
 }\\
($^{B}$) Dream \& Science Factory, Marseilles (France)\\
($^{C}$) ISEG, Universidade T\'{e}cnica de Lisboa (TULisbon) and\\ Research Unit on Complexity in Economics (UECE)\\
($^{D}$) Institute for Complexity Sciences (ICC), Lisbon (Portugal)\\

\end{small}
\end{minipage}
\end{center}

\vskip 2truecm

\begin{abstract}
This paper introduces a Markov chain approach that allows a rigorous  analysis of agent based opinion dynamics as well as other related agent based models (ABM).
By viewing the ABM dynamics as a micro description of the process, we show how the corresponding macro description is obtained by a projection construction. Then, well known conditions for lumpability make it possible to establish the cases where the macro model is still Markov. In this case we obtain a complete picture of the dynamics including the transient stage, the most interesting phase in applications. For such a purpose a crucial role is played by  the type of probability distribution used to implement the stochastic part of the model which defines the updating rule and governs the dynamics.  In addition, we show how restrictions in communication leading to the co--existence of different opinions follow from the emergence of new absorbing states.
We describe our analysis in detail with some specific models of opinion dynamics. Generalizations concerning different opinion representations as well as opinion models with other interaction mechanisms are also discussed.
We find that our method may be an attractive alternative to mean--field approaches and that this approach provides new perspectives on the modeling of opinion exchange dynamics, and more generally of other ABM.

\ms
\bs {\it Keywords}:
{\bf Agent Based Models}, {\bf Opinion Dynamics}, {\bf Markov chains},  {\bf Micro Macro}, {\bf Lumpability} , {\bf Transient Dynamics} .

\ms {\it MSC:} 37L60, 37N25, 05C69.
\end{abstract}


\section{Introduction}~\label{section-introduction}

Recent improvements in multidisciplinary methods and, particularly, the availability of powerful computational tools are giving researchers an ever greater opportunity to investigate societies in their
complex nature.
The adoption of a complex systems approach allows the modeling of macro--sociological or economic structures from a bottom--up perspective --- understood as resulting from the repeated local interaction of socio--economic agents --- without disregarding the consequences of the structures themselves on individual behavior, emergence of interaction patterns and social welfare.

\ms

Agent based models (ABM) are at the leading edge of this endeavor. When designing an agent model, one is inevitably faced with the problem of finding an acceptable compromise between realism and simplicity.
If many aspects are included into the agent description, the model might be plausible with regard to the individual behaviors, but it will be impossible to derive rigorous analytical results.
In fact, it can even be very hard to perform systematic computations to understand the model dynamics if many parameters and rules are included.
On the other hand, models that allow for an analytical treatment often oversimplify the problem at hand.
In ABM, we can find the whole spectrum between these two extremes.
While simplicity is often favored by physicists in order to be able to apply their well--developed tools from statistical physics, more realistic descriptions are often desired by researchers in the humanities because they are interested in incorporating into the model a reasonable part of their qualitative knowledge at the micro and macro scales.
Both views have, of course, their own merits.

\ms

Our paper is a contribution to interweaving two lines of research that have developed in almost separate ways: the Markov chain approach and ABMs.
The former represents the simplest form of a stochastic process while the latter puts a strong emphasis on heterogeneity and social interactions.
The main expected output of our Markov chain strategy applied to ABM is a better understanding of the relationship between microscopic and macroscopic dynamical properties.
Moreover, we aim to contribute not only to the understanding of the asymptotic properties of ABM but also to  the transient mechanisms that rule the system on intermediate time scales.
For practical purposes this is the most relevant information for two reasons: first, in our case the chains are absorbing, so the asymptotic dynamics is trivial and second,  they describe the evolution of the system before external perturbations take place and possibly throw it into a new setting.

\ms

Agent--based opinion models, a particular case of ABM, are among the most simple models in the literature and are therefore a suitable starting point for the analysis.
Especially for \emph{binary opinion models} several results have been obtained by previous authors using analytical tools, as shown in the review on social dynamics by Castellano and co-workers \cite{Castellano2009}.
The most intensively studied model is the \emph{voter model}, originally developed by Kimura and Weiss \cite{Kimura1964} as a model for spatial conflict of two species (see also Refs. \cite{Clifford1973, Frachebourg1996, Slanina2003, Sood2005, Vazquez2008, Schweitzer2008}).
The analysis of binary opinion models is usually based on mean--field arguments.
The microscopic agent configuration is mapped onto an aggregate order parameter, and the system is reformulated on the macro--scale as a differential equation which describes the temporal evolution of that parameter.

\ms

A mean--field analysis for the voter model on the complete graph was presented by Slanina and Lavicka in Ref. \cite{Slanina2003}, and naturally, we come across the same results using our method as the first part of approach.
Slanina and Lavicka derive expressions for the asymptotic exit probabilities and the mean time needed to converge, but the partial differential equations that describe the full probability distribution for the time to reach the stationary state is too difficult to be solved analytically (\cite{Slanina2003}, pag.4).
Analytical results based on the same methods have been obtained for the voter model on $d$--dimensional lattices (\cite{Cox1989,Frachebourg1996,Liggett1999,Krapivsky2003}) as well as for networks with uncorrelated degree distributions (\cite{Sood2005,Vazquez2008}).

\ms

One step to a more realistic agent description (though still a caricature) is achieved by allowing the agents to make n--ary choices.
Among the most popular models that realize this is the Axelrod model \cite{Axelrod1997} , which uses vectors as state variables and bounded confidence (\cite{Hegselmann2002,Deffuant2001}).
In both models, the interaction probability is a function of the agent similarity such that similar agents tend to interact and so become more similar in the interaction.
In the Axelrod model as well as in other bounded confidence models, this leads to the emergence of clustering as agents converge in homogeneous subgroups while an appropriate distance in between these subgroups increases. An analytical treatment of this process is already quite difficult (see Ref. \cite{Castellano2000} for an approximate mean--field analysis).
Our Markov chain approach shows how restrictions in the agent communication lead to the emergence of new absorbing states in the associated Markov chain which correspond to system states where different opinions co--exist.

\ms

The usefulness of the Markov chain formalism in the analysis of more sophisticated ABMs has been discussed by Izquierdo and co-workers (\cite{Izquierdo2009}), who look at 10 well--known social simulation models by representing them as a time--homogeneous Markov chain.
Among these models are the Schelling segregation model (\cite{Schelling1971}, for which some analytical results are available, for example, in Refs.\cite{Pollicott2001,Grauwin2010}), the Axelrod model (considered above) and the sugarscape model from Epstein and Axtell \cite{Epstein1996}.
The main idea of Izquierdo and co-workers \cite{Izquierdo2009} is to consider all possible configurations of the system as the state space of the Markov chain.
Despite the fact that all the information of the dynamics on the ABM is encoded in a Markov chain, it is difficult to learn directly from this fact, due to the huge dimension of the configuration space and its corresponding Markov transition matrix.
The work of Izquierdo and co-workers mainly relies on numerical computations to estimate the stochastic transition matrices of the models.

\ms

In our opinion, a well posed mathematical basis for these models may help the understanding of many of their observed properties.
Linking the micro--description of an ABM to a macro--description in the form of a Markov chain provides information about the transition from the interaction of individual actors to the complex macroscopic behaviors observed in social systems.
In particular, well known conditions for lumpability make it possible to decide whether the macro model is still Markov.
Conversely, this setting can also provide a suitable framework to understand the emergence of long range memory effects.

\ms

In sociology, the concepts of micro and macro have long been an important subject of analysis. Different but related meanings have been  advocated by different authors (see Ref. \cite{Alexander1987} and references therein), running from "micro as dealing with individuals and macro as dealing with populations" to "micro as social processes that engender relations among individuals and macro as the structure of different positions in a population and their constraints on interaction". In any case the terms micro and macro relate in this context the action of individuals or small groups based on their mutual relations and the emergence of collective societal scopes.

\ms

One of the first acknowledged synthetic formulations of this linkage between micro and macro in sociology studies is from Max Weber (\cite{Weber1978}, pag.29) from where we quote the following basic observation: "within the realm of social action, certain empirical uniformities can be observed, that is, courses of action that are repeated by the actor or (simultaneously) occur among numerous actors". We shall see how a stylized version of this belief is incorporated in our study when passing from micro to macro dynamics. Talcott Parsons \cite{Parsons1954} later introduced the notion of internalization, posing an interesting question on the retroaction of the macro on the micro level of the models, to which we  shall come back in section \ref{section-projected}. See Ref. \cite{Alexander1987} for a discussion of this topic.

\ms

On the other hand our work is certainly related to the social network literature, another old outstanding branch of sociology. Sociograms, an important tool in social studies, had already been introduced by Jacob Levy Moreno in the 30s, (\cite{Moreno1934}, \cite{Moreno1946}) inducing a graph representation of social relations, thus opening the way to social network based research. This approach has been developed since the seminal studies of John Barnes \cite{Barnes1954} and it expanded rapidly with the work of Mark Granovetter \cite{Granovetter1973} and many others. In our case, however, as in plenty of other models the emphasis is on the dynamics of the processes rather than on the structure alone (\cite{Helbing1994,Weidlich2002,Schweitzer2003} and references therein).

\ms

One of the key notions to understand the micro--macro linkage is emergence. We follow Ref. \cite{Giesen1987} to describe various levels of emergence present in the sociological literature and comment on the links with our work. First the so called \textit{descriptive emergence} problem: are "the macroproperties a common property of many microunits" or are they something fundamentally different? This question is clearly addressed in our work. In fact, starting from a "simple aggregative procedure" \cite{Giesen1987} on the individual attributes to define the macro level, we show how the nonlinearity features  of dynamical process may naturally create "something else" at least in a stylized model. Next, related to Parson's phenomenon of internalization \textit{practical emergence} stands for the possible discrepancy, "if the macrostructural properties of a social system no longer correspond to the internalized rules and interactions of the individual actors", which leads to "a practical problem for the individuals acting in this system". At the end of section \ref{subsection-delta2-transient} we discuss a possible quantification, in a very rough sense, of this "falling out" (\cite{Giesen1987}, pag.339) of the micro and macro levels. Finally, "far more controversial" (\cite{Giesen1987}, pag.339)  \textit{explanatory emergence} stands for a process leading to autonomous dynamics of structures (\cite{Brodbeck1968}, pags.280-303). We shall comment on this possibility in section \ref{section-generalizations}, without really answering the question, but indicating how it could be taken into account in the framework of our models.

\ms

The paper is organized as follows: in section \ref{section-microdynamics}, starting from the idea introduced by Izquierdo and co-workers \cite{Izquierdo2009} we rigorously treat a class of ABM as Markov chains .
Our main idea is to give a solid basis to the link from a micro to a macro description in the ABM context and fully explore the potential of this construction, a task to which sections \ref{section-projected} and \ref{section-opinion-projected} are devoted.
Generalizations concerning different opinion representations as well as opinion models with other interaction mechanisms are then discussed in section \ref{section-generalizations}.
We find that our method may be an attractive alternative to mean-field approaches and this approach can provide new perspectives on the modeling of opinion exchange dynamics, and more generally of other ABM.
We end up with some final remarks and a prospective for further work in section \ref{section-conclusion}.

\section{Microdynamics as a Markov chain}~\label{section-microdynamics}

\ms

Here we consider the class of ABM defined by a set N of agents, each one characterized by individual attributes that are taken in a finite list of possibilities. The meaning or the level of abstraction of the content of such attributes is not important as long as it can be codified in a finite, possibly very large, set of possibilities. The agents should also be able to regularly update their attributes according to the information conveyed from other agents as well as to their actual state. The collective updating process of the attributes of the agents at each time step consists of two parts. First a random choice of a subset of agents is made according to some probability distribution $\omega$. The number of agents chosen at each time may also be random. Then the attributes of the agents are changed (updated) according to some rule, called a dynamical rule, which depends on the the subset of agents selected at this time. We denote by $S$ the individual state or attribute space and we call the configuration space $\Sigma$ the set of all possible combination of attributes of the agents, i.e. $\Sigma = S^N$. It is noteworthy pointing out the meaning of the distribution $\omega$ in cases where an ABM is used to model human (or animal, biological, etc.) behavior. It should be thought as the idealized footprint of a collective structure, in the sense of constraints upon agent agency as well as enablers. See, for example, (see Ref. \cite{Giddens1984} for a precise sense of structure as used in this context). Notice that, at this level, the dynamics of the model is defined in the configuration space, which seeks to describe the dynamics of each agent in full detail. We shall refer to this as micro dynamics.

\ms

Let $\Sigma $ be a finite set (ex. the configuration space of an ABM) and $Z$ index a collection of maps $\{F_z, z\in Z\}, F_z : \Sigma \rightarrow   \Sigma$ and $\omega$ a probability distribution on $Z$.
If $F_{z_1}, F_{z_2}, ...$ is a sequence of independent random maps, each having distribution $\omega$, and $X_0 \in \Sigma$ has distribution $\mu_0$, then the sequence $X_0, X_1, ...$ defined by
\begin{equation}
X_t= F_{z_t}  (X_{t-1}), t \geq 1
\label{RMR1}
\end{equation}
is a Markov chain on $\Sigma$ with transition matrix $\hat{P}$:
\begin{equation}\label{RMR2}
\hat{P}(x,y)  = {\bf{Pr}_{\omega}}(z \in Z, F_z (x) =  y), x,y \in \Sigma
\end{equation}

\ms
Conversely (\cite{Levin2009}), any Markov chain has a random map representation (RMR). Therefore (\ref{RMR1}) and  (\ref{RMR2}) may be taken as an equivalent definition of a Markov chain which is particularly useful in our case, because it shows that an ABM that can be described as above is, from a mathematical point of view, a Markov chain and can be treated following the approach developed in this work. This class includes the models described in Ref. \cite{Izquierdo2009}.
\ms

\ms

In order to motivate and exemplify our point of view we introduce a model of opinion dynamics that is a particular case of the class of ABM described above.
We  shall come back to this model during the paper to illustrate our results and to show how they lead up to final conclusions in this particular case.

\ms

To define the opinion model, let us consider a population $\bf{N}$ of $N$ agents and denote the opinion, which is the single attribute of agent $i$ at time $t$ as $x_i(t) \in S$, where we assume that each agent can choose out of $\delta$ different alternative opinions so that the agent state space $S = \left\{0,1,\ldots,\delta-1\right\}$ has $\delta$ states.
Let $x(t) = \{x_1(t),\ldots,x_N(t)\}$ describe the opinions of all the agents at time $t$.
We refer to this as the \emph{opinion configuration} or the \emph{opinion profile} of the population.
Then the \emph{configuration space} of the model is $\Sigma = S^{\bf{N}}$, the space of all possible configurations $x$.

\ms

In this very simple opinion model  the willingness of two agents ($i,j$) to communicate depends on the similarity of their opinions.
This can be encoded in form of a confidence matrix $\alpha: S\times S \rightarrow \left\{0,1\right\}$ such that $\alpha(s_1,s_2) = 1$ if the opinions $s_1$ and $s_2$ are sympathetic $\alpha(s_1,s_2) = 0$ if they are not.\footnote{The reason to call $\alpha$ the confidence matrix came from the fact that we can define this matrix by means of a confidence threshold, usually defined by means of a threshold on a distance on the set $S$ of the possible opinions. The confidence matrix is defined, for any pair (x,y) of elements of  $\bf{S}$ by $\alpha (x,y) = 0$ if the relative confidence $(x,y)$ is below threshold and $\alpha (x,y) = 1$ if not. Notice that a confidence matrix is a generalization of the notion of confidence threshold.}
In the iteration process of this model, an agent pair $(i,j) \in {\bf{N}}\times {\bf{N}}$ is chosen at random "to meet" according to a probability distribution $\omega$, ex. $\omega$ is the uniform distribution, $\omega(i,j) = \frac{1}{N^2}$, for all $i,j$.
If two agents $(i,j)$ meet and $\alpha(x_i(t),x_j(t)) = 1$ then agent $i$ imitates agent $j$, that is $x_i(t+1) = x_j(t)$, which brings us to define an updating function by $ \mathbf{u}  (x_i,x_j)= (x_j, x_j)$. If, instead,
$\alpha(x_i(t),x_j(t)) = 0$ the opinion configuration is not changed. Because in this version of the model the first chosen agent always imitates the second, the model is sometimes called a directed opinion model. \ms

\ms

In fact we treat a more general class of ABM with the following characteristics:

\ms

(H1) There is a finite number $N$ of agents. The set of all agents is denoted by $\bf{N}$. Each agent is characterized by a single attribute which takes a value among $ \delta $ possible alternatives. The set of  $ \delta $ elements of possible situations is denoted by $\bf{S}$. The configuration space is the set $\bf{\Sigma} = \bf{S^N}$ of all possible combinations of the situations for the $N$ agents. Therefore, if $\vec{x} \in \bf{\Sigma}$ we write $\vec{x} = (x_1,\dots, x_i, \dots, x_N)$ with $x_i \in \bf{S}$.

\ms

(H2) There is a $\delta \times \delta $ matrix $\alpha$ with entries $0$ or $1$ called the \textbf{confidence matrix}  and an updating function denoted by $\bf{u}$ that is a function from $\bf{S} \times \bf{S}$ into $\bf{S} \times \bf{S}$.

\ms

(H3) The ABM is defined as a discrete time Random Walk in the configuration space $\bf{\Sigma}$. If the walker is in a configuration $\vec{x}$ at time $t$, it will jump to the configuration $\vec{y}$ under the following prescription:
\begin{enumerate}
\item
Two agents\footnote{In some ABM the updating is made by choosing a number of agents different from 2, or even different at each time step. As is shown section ~\ref{section-generalizations}, it is easy to generalize what follows in this case by just defining $\omega$, $\alpha$ and $\bf{u}$ accordingly.} $i$ and $j$ are chosen according to a given probability distribution $\omega$ on $\bf{N} \times \bf{N}$.
\item
Compute $\alpha (x_i,x_j)$. If $\alpha (x_i,x_j) = 0$ the walker doesn't move. If $\alpha (x_i,x_j) = 1$ then the walker moves to a configuration $\vec{y} = (x_1,\dots, x_{i-1}, y_i, x_{i+1} \dots, x_{j-1}, y_j, x_{j+1}, \dots, x_N)$ which defers from $\vec{x}$ only eventually in the value of the attributes of the agents $i$ and $j$.
\end{enumerate}

\ms
That is:

\begin{equation}\label{update_function}
(y_i,y_j) = (x_i,x_j) \ \mbox{if} \ \alpha (x_i,x_j) = 0 \nonumber \\
\end{equation}
and
\begin{equation}\label{update_function_bis}
 (y_i,y_j)= \mathbf{u}  (x_i,x_j) \ \mbox{if} \ \alpha (x_i,x_j) = 1\\
\end{equation}

\ms
\begin{note}\label{no-alpha}

The use of a confidence matrix $\alpha$ is just a matter of convenience. It would be possible to absorb $\alpha$ in $\bf{u}$ by forcing $  \mathbf{u}  (x_i,x_j) = (x_i,x_j) \ if \ \alpha (x_i,x_j) = 0$. We use this double encoding of the updating rule to follow the tradition in the field, assigning the updating rule $\bf{u}$ to "the model" and the confidence matrix $\alpha$ to "a parameter of the model".
\end{note}

\ms

As is clear from the previous discussion, under the hypotheses (H1), H(2), (H3) the ABM is an homogeneous Markov chain with state space $\bf{\Sigma}$ being the configuration space and a transition probability $\hat{P}(\vec{x},\vec{y})$ as defined below.

\ms

Notice that, according to the transition rule described above, the possible jumps are constrained to a ball  of "Hamming" radius 2 around the actual configuration.

\ms

We describe now the transition probability matrix of the Markov chain according to Eq. (\ref{RMR2}).
Let us say that a pair of configurations $\vec{x}$ and $\vec{y}$ are \textbf{adjacent} if all the agents have the same attribute values in x and in y except possibly the agents $i$ and $j$, i.e. $x_k = y_k$ if  $k \neq i$ and $k \neq j$. These are denoted $\vec{x} \stackrel{i,j}{\sim} \vec{y}$.
Then we define a transition probability matrix $\hat{P}$ of a Markov chain by

\ms

\begin{equation}\label{Phat1}
\mathrm{if} \: \vec{x} \stackrel{i,j}{\sim} \vec{y} \:\mathrm{and} \: x \neq y\\\: \mathrm{then}\:
\hat{P}(\vec{x},\vec{y}) = \sum_{\substack{(x_i,x_j), i\neq j, \\ (y_i,y_j) = \mathbf{u}  (x_i,x_j)\\ \alpha(x_i,x_j)=1}} \omega(i,j)
\end{equation}

\ms
\begin{equation}\label{Phat2}
\hat{P}(\vec{x},\vec{x}) = 1 - \kappa(\vec{x}) \\
\end{equation}

\ms
\begin{equation}\label{Phat3}
\hat{P}(\vec{x},\vec{y}) = 0  \: \mathrm{elsewhere} \\,
\end{equation}

\ms
where
\begin{equation}\label{kappa}
\kappa(\vec{x}) =  \sum^{}_{\substack{\vec{y} {\sim} \vec{x}}} \\ \sum_{\substack{ i\neq j, \\ (y_i,y_j) = \mathbf{u}  (x_i,x_j)\\ \alpha(x_i,x_j)=1}} \omega(i,j)
\end{equation}

\ms
 Notice that the summation in (\ref{Phat1}) is needed because, in general, the updating function is not one to one in the configuration space so several links may join two given adjacent configurations. We shall see an example of such a situation in the opinion model.

\ms

The transition probability matrix of a Markov chain is of dimension $\bf{S^N} \times \bf{S^N}$ and it describes the evolution of any initial distribution on the configuration (state) space.
The comparison with a numerical simulation starting from some particular initial configuration may therefore be performed using the corresponding Dirac distribution. But, even in this case, the Markov analysis will spread this initial concentrated distribution in the course of the updating process. Clearly it is possible to use other initial distributions for the Markov chain, and each distribution will correspond to a set of numerical simulations with different initial configurations chosen according to the distribution.

\ms

Each row of the transition probability matrix has at most $N^2$ non--zero elements (including a loop).
This is due to the adjacency criteria, which encodes the fact that only a pair of agents may change their state from one step to the other.
In fact, the non zero elements inside the row $\vec{x}$ are determined by the possible choices of couples $x_i,x_j$ of components (agents) of  $\vec{x}$. These non--zero elements will appear in the columns corresponding to the $\vec{y}$ adjacent to $\vec{x}$ for which $\alpha (x_i,x_j) = 1$.

\ms

Despite the fact that all the information of the dynamics of the ABM is encoded in such a Markov chain, it is not always easy to learn directly from this fact, because of the over large dimension of the configuration space and its corresponding Markov transition matrix. As an example of a question that may be answered at this level, we mention the characterization of absorbing configurations. These are the $\vec{x} = (x_1,\dots, x_i, \dots, x_N)$ such that, for any $(x_i,x_j)$, $\alpha (x_i,x_j) = 0$ or $ \mathbf{u}  (x_i,x_j)= (x_i, x_j)$ and they are easy to identify even in the case of bounded confidence models.  It is clear that the properties of the transient dynamics are most relevant for Markov chains with absorbing configurations. The analytical study of these properties is based upon the corresponding fundamental matrix \cite{Kemeny1976}. But such computation needs to invert a matrix of order $\mid S^N \mid$ that can be performed only numerically. Therefore another strategy is lacking to go further in this direction. In the next section we shall present one possibility to overcome this difficulty.

\ms

As already noticed, the directed opinion model is a special case of such a formalism.
There, only one coordinate of the configuration may change at each step.  The transition probability matrix $\hat{P}$ is obtained from (\ref{Phat1}), (\ref{Phat2}), (\ref{Phat3}) and(\ref{kappa}) by restricting the corresponding sums to the first index $i$ only.

\ms

One immediate consequence of the topology of this transition matrix is that in the opinion model the only absorbing configurations are complete consensus (all the agents having the same opinion) or mutually antagonistic (agents having opinions $s$ and $s^{'})$ for which $\alpha (s,s^{'}) = 0$ for all pairs $(s,s^{'})$). Recall that, with probability one, the system falls in one of the absorbing states in finite (although not uniform) time. Therefore we see how in this model the lack of confidence allows for new absorbing states, stabilizing non--consensual opinion profiles.

\section{Macrodynamics, Projected Systems and Observables.}~\label{section-projected}

\ms

A projection of a Markov chain with state space $\bf{\Sigma}$ is defined by a new state space $\bf{X}$
and a (projection) map $\Pi$ from $\bf{\Sigma}$ to $\bf{X}$.
The meaning of the projection $\Pi$ is to lump sets of micro configurations in $\bf{\Sigma}$ accordingly to some macro property in such a way that, for each $X \in \bf{X}$, all the configurations of $\bf{\Sigma}$ in $\Pi^{-1} (X)$ share the same property.

 \ms
 
 In fact such projections are important when catching the "macroscopic" properties of the corresponding ABM because they are in complete correspondence with a classification based on an observable property of the system. To see how this correspondence works let us suppose  that we are interested in some factual property of our agent based system. This means that we are able to assign to each configuration the specific value of its corresponding property. Regardless of the kind of value used to specify the property(qualitative or quantitative), the set $\bf{X}$ needed to describe the configurations with respect to the given property is a finite set, because the set of all configurations is also finite. Let then $\phi : \bf{\Sigma} \rightarrow \bf{X} $  be the function that assigns to any configuration $x \in \bf{\Sigma}$ the corresponding value of the considered property. It is natural to call such $\phi$ an observable of the system. Now, any observable of the system naturally defines a projection $\Pi$ by lumping the set of all the configurations with the same $\phi$ value. Conversely any  (projection) map $\Pi$ from $\bf{\Sigma}$ to $\bf{X}$ defines an observable $\phi$ with values in the image set $\bf{X}$. Therefore these two ways of describing the construction of a macrodynamics are equivalent and the choice of one or the other point of view is just a matter of taste.

 \ms

The price to pay in passing from the micro to the macrodynamics in this sense (\cite{Kemeny1976}, \cite{Chazottes2003}) is that the projected system is, in general,  no longer a Markov chain:  long memory (even infinite) may appear in the projected system.

\ms

Given the transition probability and the initial distribution defining the first Markov chain in $\bf{\Sigma}$, we are interested in the projected measure $\mu$ defined for all cylinders $[X_{(1)},X_{(2)}, \dots, X_{(r)}]$ of  $\bf{X}$ by:

\ms
\begin{equation}\label{markov_projection}
\mu [X_{(1)},X_{(2)}, \dots, X_{(r)}] = \hat\mu \Pi^{-1}  [X_{(1)},X_{(2)}, \dots, X_{(r)}]
\end{equation}

\ms
 where $\hat\mu$ denotes the corresponding probability for the  initial Markov chain.

\ms

The conditions for a projection of a Markov chain still to be a Markov chain are known as lumpability (or strong lumpability),  and necessary and sufficient conditions for lumpability are known \cite{Kemeny1976}.
 In general it may happen that, for a given Markov chain, some projections are Markov and others not. Therefore a  judicious choice of the macro properties to be studied may help the analysis.
In order to establish the lumpability in the cases of interest we shall use symmetries of the model. For further convenience, we  state a result for which the proof is easily given Thm. 6.3.2 of  \cite{Kemeny1976}:

\begin{proposition}\label{proposition_lumpability}
\ms
Let  $(\bf{\Sigma}, \hat{P}))$ be a Markov chain and $(X_1,\dots, X_{n})$ a partition of $\bf{\Sigma}$.
Suppose that there exists a set ${\Lambda}$ of bijections of $\bf{\Sigma}$ (therefore a group of symmetries ) such that:

\ms

(1) ${\Lambda}$ preserves the partition: for any ${\lambda} \in \Lambda $ and any  atom $X_j$, we have ${\lambda} X_{j} \subseteq X_{j}$.

\ms

(2)   ${\Lambda}$ acts transitively on each $X_{j}: X_{j} = \bigcup _{\lambda} {\lambda}x$,  for some (and then all) $x\in X_j$.

\ms

(3) The Markov transition probability $\hat{P}$ is symmetric with respect to $\Lambda$:

\begin{equation}\label{symmetry_lumpability}
 {\hat P} (x,y) = \: {\hat P} ({\lambda}x,{\lambda}y) \: \mathrm{for \: any}  {\lambda} \in {\Lambda}
\end{equation}

\ms
then the partition  $(X_1,\dots, X_{n})$ is (strongly) lumpable.
 \end{proposition}

\ms
The opinion model is a nice example where such a projection construction is particularly meaningful. There, we consider the projection $\Pi$ that maps each $\vec{x} \in\bf{\Sigma} $ into $X_{<k_0,\dots,k_{\delta}-1>} \in \bf{X}$ where $k_s$, $s=1,\dots, {\delta}$, is the number of agents in $\vec{x}$  with opinion $s$. The projected configuration space is then made of the $X_{<k_0,\dots,k_{\delta}-1>}$ where $k_s \geq 0$, $s=1,\dots, {\delta}-1$ and $\sum_{0}^{{\delta}-1} k_{s} = N$.

 \ms
 \section{Opinion Dynamics and Projected Systems}~\label{section-opinion-projected}

 \ms
 
 We shall now treat in detail the opinion model as an example of the previous ideas.

 \ms

 \subsection{The Macro Dynamics of Binary Opinion Model}~\label{subsection-delta2}

\ms

The case of a binary opinion model, $\delta=2$, is particularly simple and therefore well-suited for an analytical starting point. In this case bounded confidence is excluded. In binary state opinion models, the opinion of agent $i$ at time $t$ is a binary variable $x_i(t) \in \left\{0,1\right\}$. The opinion profile is given by the bit--string $x(t) = \{x_1(t),\ldots,x_N(t)\}$.  The space of all possible configurations is  $\Sigma = \left\{0,1\right\}^N$.

\ms

Let us define a function $N_1$ on the configuration space $\Sigma$ such that
\begin{equation}
N_1(x) = \sum_{i=1}^N x_i, \forall x \in \Sigma.
\label{eq:F1T2.N1}
\end{equation}
It counts the number of agents in state $1$.
Using $N_1(x)$, we define $X_k \subset \Sigma$ by

\begin{equation}
X_{k} = \left\{x : N_1(x) = k \right\}.
\label{eq:F1T2.Xk}
\end{equation}

\ms
Each $X_k \subset X, k = 0 \ldots N$ contains all the configurations ($x$) in which exactly $k$ agents hold opinion $1$ (and then $N-k$ hold opinion $0$).
In this way we obtain a partition of the configuration space $\Sigma$.
Notice that $X_0$ and $X_N$ contain only one configuration, namely $X_0 = \left\{\vec{0}\right\}$ and $X_N = \left\{\vec{1}\right\}$.

\ms

Using the group $\G_N$ of all the permutations of $N$ agents, it is clear that such a partition fulfills conditions (1) and (2) of Proposition(\ref{proposition_lumpability}).
So lumpability of this partition leans on condition (3) of Proposition(\ref{proposition_lumpability}): the invariance of the Markov transition matrix $\hat P$ under the permutation group of agents. Notice that no restriction on the confidence matrix is needed for it only depends on the opinions and not on the labeling of the agents. In fact, the probability distribution $\omega$ must be invariant under the permutation group  $\G_N$ and therefore uniform:  $\omega(i,j) = \frac{1}{N^2}$, for all pair of agents ($i,j$).

\ms

It is worthwhile noticing at this point that the uniform distribution, corresponding to the most unstructured dynamical rule, still entails emergent   organized patterns in the system of opinions. Because the set of opinions is dynamically organized, the homogeneity of the uniform distribution on the agent population has an implicit structure when viewed through the opinion content.

\ms

Moreover, for some other distributions $\omega$, it may be possible to refine the partition so as to get lumpability. For instance, if the agents are divided in subsets in which $\omega$ is constant, then the partition defined  by an equal number of opinions inside each subclass is lumpable. In this case $\alpha$ will depend on the pair labeling of agents together with their respective opinions and not only on the latter. The block structure of $\alpha$ then determines the projection scheme.

\ms

For the model with complete confidence, $\alpha (s,s^{'}) = 1$ for any $(s,s^{'})$, and uniform distribution $\omega$, the Markov chain is defined by the stochastic transition matrix:

\begin{equation}
\renewcommand\arraystretch{1.25}
P =
	\left( {\begin{array}{*{20}c}
   {1 } & {0}  & {0} & {0 } & {0} & {\ldots} & {0}  \\
   {p(1)} & {q(1)} & p(1) & {0 } & {0} & {\ldots} & {0}  \\
   {0}  & {p(2)} & {q(2)} & p(2) & {0} &  {\ldots } & {0}   \\

   {\vdots} & {\ddots} & {\ddots} &  {\ddots} & {} &  {} & {\vdots}  \\
   { 0 } & {} & {p(k)} & {q(k)} & {p(k)} & { } & {0}  \\
   {\vdots} & {} & {\ddots} & {\ddots} & {\ddots} &  {} & {\vdots} \\
   {0} & {\ldots } & {0} &{p(N-2)} & {q(N-2)} & {p(N-2)} & {0}  \\
   {0} & {\ldots } & {0} & {0} &{p(N-1)} & {q(N-1)} & {p(N-1)}  \\
   {0} &  {\ldots} & {0} & {0}  & {0} &  {0} & {1}  \\
\end{array}} \right),
\label{eq:PMat}
\end{equation}

\ms
with

\ms
\begin{eqnarray}\label{eq:F1T2.p}
p(k) = \frac{k (N-k)}{N^2}.
\end{eqnarray}

\ms
and

\ms
\begin{eqnarray}\label{eq:F1T2.q}
q(k) = 1- 2 p(k) =\frac{k^2 + (N-k)^2}{N^2}.
\end{eqnarray}

Formulas (\ref{eq:F1T2.p}) and(\ref{eq:F1T2.q}) follow from Thm. 6.3.2 of \cite{Kemeny1976}  and the fact that for any micro--configuration in $X_k$ there are $k (N-k)$ possible meetings taking it to $X_{k+1}$ and the same number of possible encounters taking it to  $X_{k-1}$.
Notice that in this case $X_0$ and $X_N$ are the only absorbing states of the process.

\ms

The probability that any opinion change happens in the system is $2p(k)$ and then depends on the current opinion balance. But there is no general tendency of the system to be attracted by one of the extremes.
Due to the particular form of $p(k)$ the prevalence of one opinion results in a reduced probability of further opinion change, contrary to the usual random walk with constant transition probabilities.

\ms

For $k \cong \frac{N}{2}$ we have $p(k) \cong \frac{1}{4}$.
By contrast, when $k$ is closed to $0$ or $N$, there is a large probability for the system to stay unchanged. Notice that for $k=1$ or $k=N-1$ this probability tends to $1$ when $N\rightarrow \infty$.
This indicates that in this model once one opinion dominates over the other, public opinion as a whole becomes less dynamic, which also reveals a difficulty for new opinions to spread in the artificial society.

 \ms
  \subsection{Transient in the Macro Dynamics of Binary Opinion Model}~\label{subsection-delta2-transient}

 \ms
 
In Markov chains with absorbing states (and therefore in ABM) the asymptotic status is quite trivial. As a result, it is the understanding of the transient that becomes the interesting issue.
We shall now analyze the transient dynamics for the macro dynamics of the binary opinion model. In order to do so, all that is needed is to compute the fundamental matrix $\bf{F}$ of the Markov chain (\cite{Kemeny1976, Behrends2000}).

\ms

Let us express $P$ in its standard form in which the two first rows and columns stand for the absorbing states $X_0$ and $X_N$ and the remaining for the $N-1$ transient states:
\begin{equation}
\renewcommand\arraystretch{1.25}
P=
	\left( {\begin{array}{*{20}c}
  {1} & {|} & { 0} \\\hline
  { R} & {|} & { Q}\\
\end{array}} \right).
\label{eq:CanonicalForm}
\end{equation}

Here  $Q$ is the $(N-1) \times (N-1)$ matrix corresponding to the transient states (without the first two rows and columns associated with $X_0$ and $X_N$).
The fundamental matrix $F$ is the inverse of $(\bf{1} - Q)$ where $\bf{1}$ is the $(N-1) \times (N-1)$ identity matrix.
Due to the structure of $P$, $(\bf{1} - Q)$ is a tridiagonal matrix that can be inverted easily using for instance the tridiagonal matrix algorithm (also known as Thomas algorithm \cite{Conte1980}).

\ms

We get:
\begin{eqnarray}\label{eq:Fij1}
 F_{ij} = \frac{N(N-i)}{N-j} : i \geq j
\end{eqnarray}

\ms
and
	
\begin{eqnarray}\label{eq:Fij2}
F_{ij} = \frac{Ni}{j} : i \leq j
\end{eqnarray}

\ms
Equations (\ref{eq:Fij1}), (\ref{eq:Fij2}) provide us with the fundamental matrix of the system for an arbitrary number of agents $N$, giving information about mean quantities of the transient dynamics in this model.

\ms

The corresponding matrix $\bf{G}$ that encodes information about the variance \cite{Kemeny1976} of the same quantities reads:

\ms

\begin{eqnarray}\label{eq:Gij1}
 G_{ij} =  (2N^2 -N) \frac{(N-i)}{(N-j)} - N^2 \frac{(N-i)^2}{(N-j)^2}: i > j
\end{eqnarray}

\ms

\begin{eqnarray}\label{eq:Gij2}
 G_{ii} =  N (N-1)
\end{eqnarray}

\ms
and
	
\begin{eqnarray}\label{eq:Gij3}
G_{ij} =  (2N^2 -N) \frac{i}{j} - N^2 \frac{i^2}{j^2}: i < j
\end{eqnarray}

\ms

An interesting quantity to characterize opinion dynamics is the time a process starting in $X_k$ takes to end in one of the two consensual absorbing states. Defining $\tau_k$ and $\upsilon_k$ as the mean and the variance of the random variable for $k= 1,\dots, N-1$ we got from (\ref{eq:Fij1}, \ref{eq:Fij2}) and \cite{Kemeny1976}:

\ms
\begin{equation}\label{eq:tauk}
	\tau_k = N \left[  \sum^{k-1}_{j=1} \frac{(N-k)}{(N-j)}  + 1 + \sum^{N-1}_{j= k+1} \frac{k }{j}\right]
\end{equation}
and the corresponding expression for $\upsilon $ can explicitly be written from (\ref{eq:Gij1}, \ref{eq:Gij2}, \ref{eq:Gij3}) using:

\ms
\begin{equation}\label{eq:tupsilon}
\upsilon = (2 {\bf F} - \bf{1}) \tau - \tau_{\it{sq}}
\end{equation}

\ms
where $\tau_{\it{sq}}$ denotes the vector resulting from $\tau$ by squaring each entry.

\ms

Therefore

\ms
\begin{eqnarray}\label{eq:tupsilonk}
\upsilon_k =  2N^2 (N-k)\left[ \sum^{k-1}_{i=1} \frac{1}{(N-i)} \left( \sum^{i-1}_{j=1} \frac{(N-i)}{(N-j)} + 1 +  \sum^{N-1}_{j=i+1} \frac{i} {j} \right) \right] +\\\nonumber
+ (2N-1) N \left( \sum^{k-1}_{j=1} \frac{(N-k)} {(N-j)} + 1 + \sum^{N-1}_{j=k+1} \frac{k}{j} \right) +\\\nonumber
 + 2 N^2 k \left[ \sum^{N-1}_{i=1} \frac{1}{k+i} \left( \sum^{k+i-1}_{j=1} \frac{(N-k-i)} {(N-j)} + 1 +  \sum^{N-1}_{j=k+i+1} \frac{k+i} {j} \right) -\right] -\\\nonumber
- N^2 \left( \sum^{k-1}_{j=1} \frac{(N-k)}{N-j} + 1 + \sum^{N-1}_{j= k+1} \frac{k}{j}\right)^2
\end{eqnarray}

\ms

\begin{figure}[htbp]
	\centering
		\includegraphics[width=0.60\textwidth]{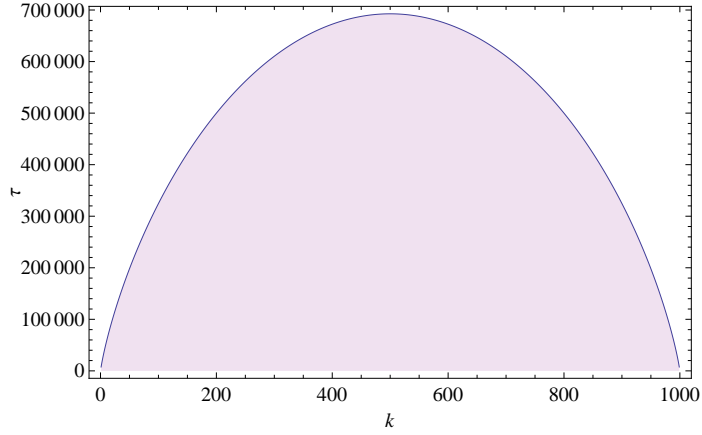}
	\caption{Mean time $\tau_k$ until absorption as a function of the initial configuration $k$ for $N=1000$. }
	\label{fig:tau1000}
\end{figure}

\begin{figure}[htbp]
	\centering
		\includegraphics[width=0.60\textwidth]{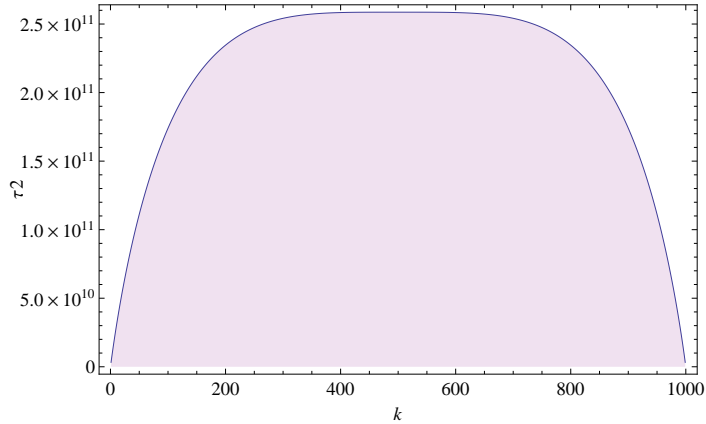}
	\caption{Variance $\upsilon_k$ until absorption as a function of the initial configuration $X_k$ for $N=1000$.}
	\label{fig:tau21000}
\end{figure}

For a system of 1000 agents, Fig.~\ref{fig:tau1000} shows the mean times until absorption $\tau_k$ from each $X_k$ and Fig.~\ref{fig:tau21000} the corresponding variances $\upsilon_k $. Notice the contrast among the two scales showing how the variance is large compared with the mean.

There are interesting consequences of (\ref{eq:tauk}) and (\ref{eq:tupsilonk}), in cases where the number of agents ($N$) becomes large. First, as already pointed out, we see that the ratio between the variance and the mean is quite large and in fact it diverges with $N$. Hence, the means are fairly unreliable estimates in this system. This is often the case for absorbing Markov chains \cite{Kemeny1976} making a direct interpretation of numerical simulations for this type of models tough. Even more subtle, the time scale depends significantly on the starting configuration $k$. In fact $\tau_k$ scales as $N \log N$ for $k=1$ and $k= N-1$ but as $N^2$ for $k = \frac{N}{2}$. We are therefore faced with a situation where to take the limit of asymptotic times first and then large number of agents or to do it in the reverse order is not equivalent. In other words, for a finite, even large, number of agents, there is a probability 1 of reaching one of the consensual configurations in finite time. By contrast, in the limit of an infinite number of agents this probability is 0 and the process will stay essentially in the configurations close to parity, $k = \frac{N}{2}$. Together with the presence of large fluctuations revealed in (\ref{eq:tupsilonk}) this fact is the imprint of a (dynamical) phase transition.

\ms

\begin{figure}[htbp]
	\centering
	 \includegraphics[width=0.60\textwidth]{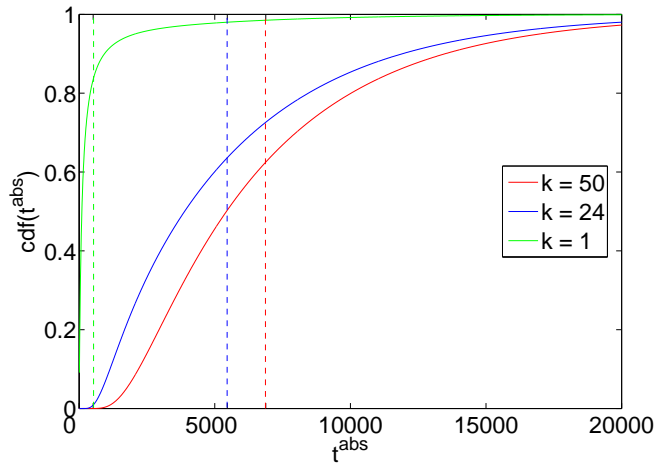}
	\caption{Cumulative probability of being absorbed  after time $t^{abs}$ (r.h.s.)  for $N = 100$ and three starting configurations $k = 1$ (green), $k = 24$ (blue) and $k = 50$ (red). Vertical lines show the respective expected absorbing times $\tau$.}
	\label{fig:absorbing_times_cdf}
\end{figure}

\begin{figure}[htbp]
	\centering
	 \includegraphics[width=0.60\textwidth]{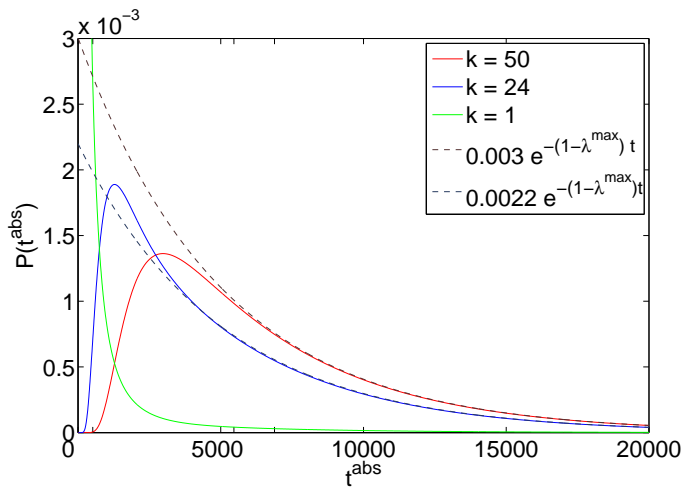}
	\caption{Probability of absorbency at time $t^{abs}$ for $N = 100$ and three starting configurations $k = 1$ (green), $k = 24$ (blue) and $k = 50$ (red). Exponential functions (dashed) are shown to illustrate the exponential decay of the convergence times.}
	\label{fig:absorbing_times_p}
\end{figure}

Besides this analysis of the scaling law of the dynamics for large $N$, it is also interesting to have an insight into the distributions of absorbing times for a system of fixed number of agents, the second item mentioned above.
As known by the Perron--Frobenius Theorem \cite{Seneta2006} this distribution is exponential  for large $t$ with rate ($1-\lambda_{max}$), $\lambda_{max}$ being the maximal eigenvalue of the matrix $Q$.
However, the correction to this distribution for intermediate times depends on the initial configuration. Indeed in our case, the distribution of the times taken by the process to fall into one of the consensual configurations departs from the exponential in a way that is strongly dependent upon the initial state, as shown in  Figs.~\ref{fig:absorbing_times_cdf} and ~\ref{fig:absorbing_times_p}.

\ms

The computation of the full time distribution is based on the fact that the powers $Q^t$ of $Q$ contain all the information about the probability that the process is still not absorbed after $t$ steps.
To be precise, the sum over the $k$th row of $Q^t$ equals the probability that the process starting at $X_k$ is not absorbed after $t$ iterations.
This yields the cumulative distribution function shown in Fig.~\ref{fig:absorbing_times_cdf} for a system of 100 agents and three starting configurations  $k = 1$ (green), $k = 24$ (blue) and $k = 50$ (red).
The vertical dashed lines represent the respective mean values $\tau$ obtained using Eq.\ref{eq:tauk}.
For $k = 50$ it becomes clear that around 60\% of simulation runs are absorbed until the expected absorption time is reached.
Fig.~\ref{fig:absorbing_times_p} shows the probability that the process is absorbed exactly at time $t^{abs}$.
The three solid curves represent the respective probabilities for $k = 1, 24, 50$.
The dashed curves are exponential functions that fit the distributions for large $t^{abs}$ showing that the distributions decay with ($1-\lambda_{max}$) as claimed above.

\ms

This leads to an interesting feature of the distribution of the absorption times coming from the fact that $\lambda_{max}$ tends to one when $N \rightarrow \infty$.
More precisely \cite{Seneta2006} (\ref{eq:PMat}) implies
\begin{equation}\label{eq:lamda_max}
1 > \lambda_{max} \geq 1 - p(1) \geq \frac{N-1}{N}.
\end{equation}
 As a consequence, we see that the times for the system to get absorbed in the final states diverge with $N$, and $Q$ approaches a stochastic matrix. In fact in the limit of infinite $N$ consensus cannot be reached. This is not the only reason why the dynamics inside the transient configurations is so important. In fact we might speculate that, in a more realistic description, exogenous events may interfere with the system and reset it from time to time, and then, in view of the previous analysis, even when the number of agents is finite but sufficiently large, the system will similarly never fall into a final absorbing unanimity configuration.

\ms

\begin{figure}[htbp]
	\centering
	\begin{tabular}{c c c}
\includegraphics[width=0.3\textwidth]{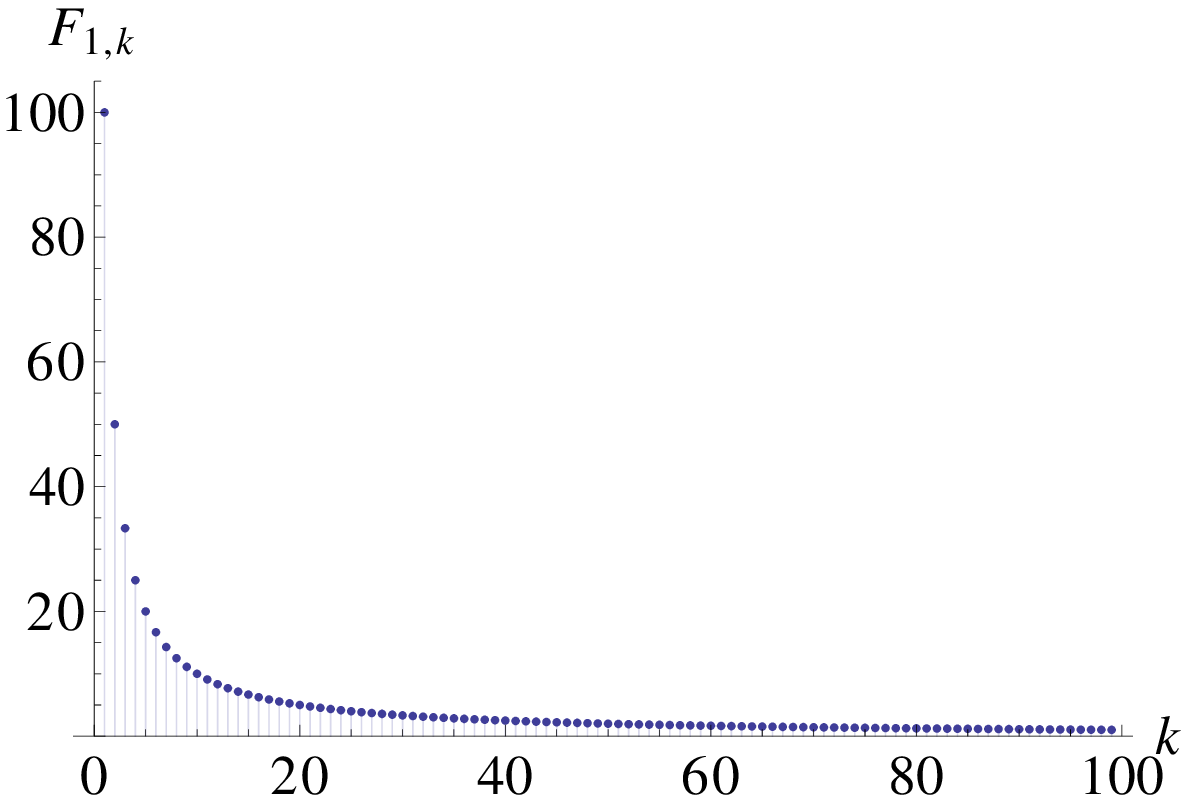}&\includegraphics[width=0.3\textwidth]{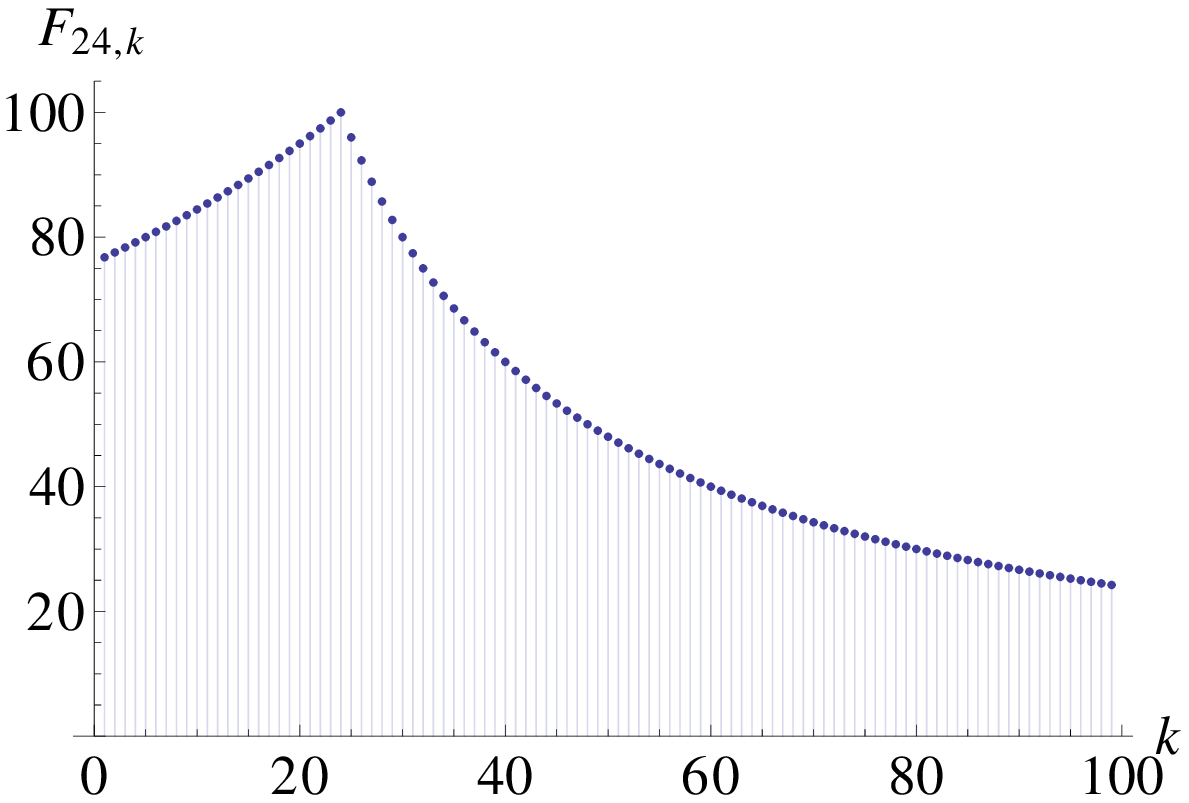}&\includegraphics[width=0.3\textwidth]{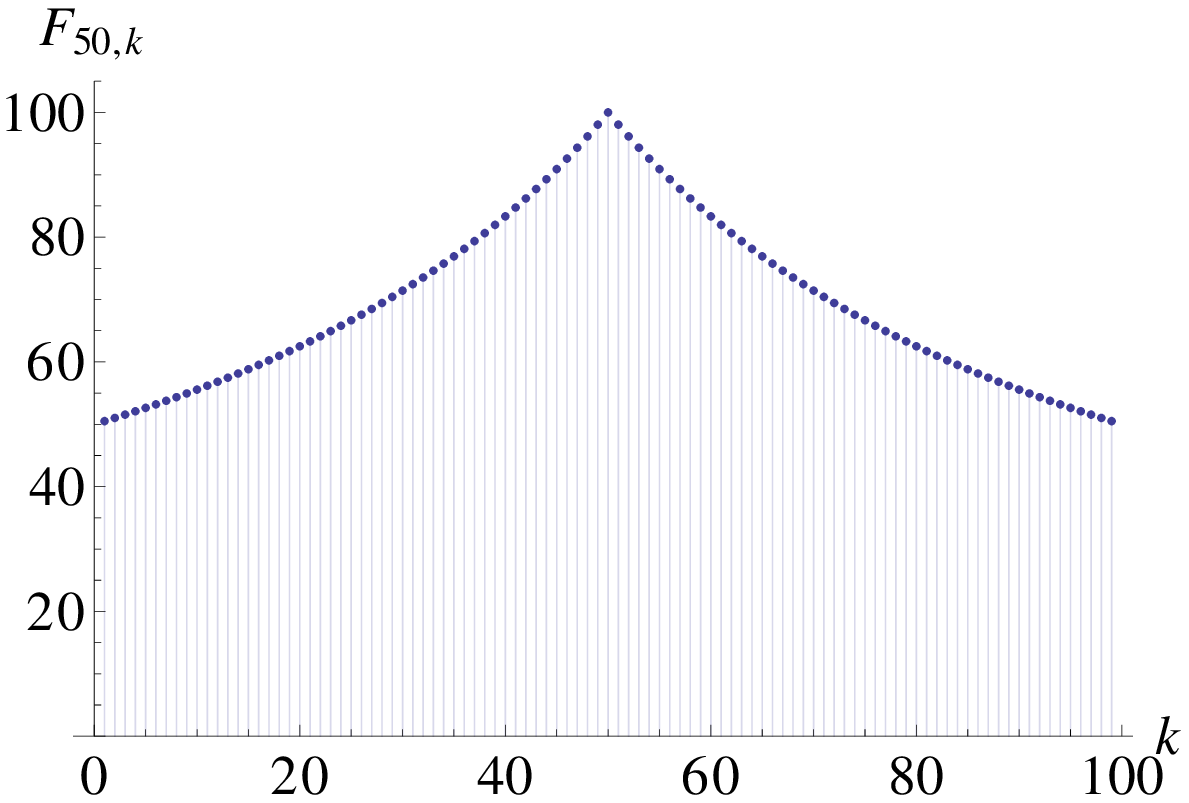}
	\end{tabular}
	\caption{The mean times for the process in a configuration $X_k$ before absorption for a walk starting in $X_1, X_{24}$ and $X_{50}$ as function of $k$ for $N=100$.}
	\label{fig:F_DifferentJ}
\end{figure}

\begin{figure}[htbp]
	\centering
	\begin{tabular}{c c c}
\includegraphics[width=0.3\textwidth]{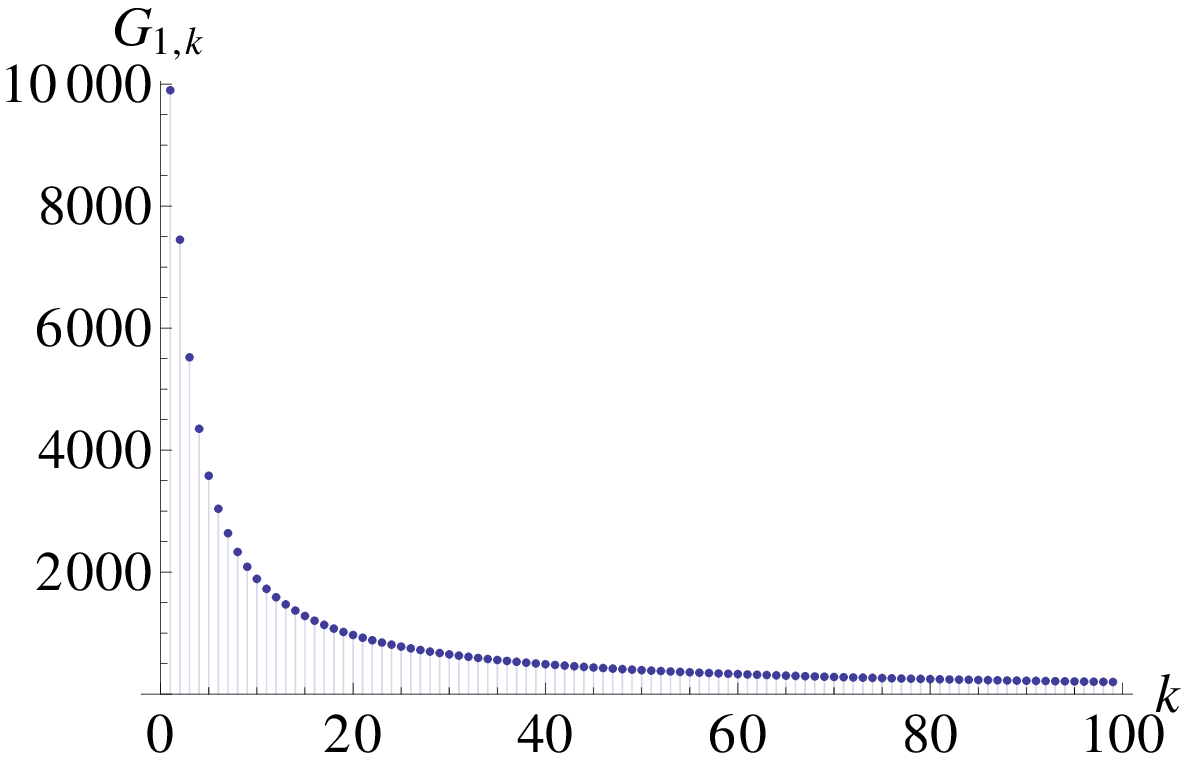}&\includegraphics[width=0.3\textwidth]{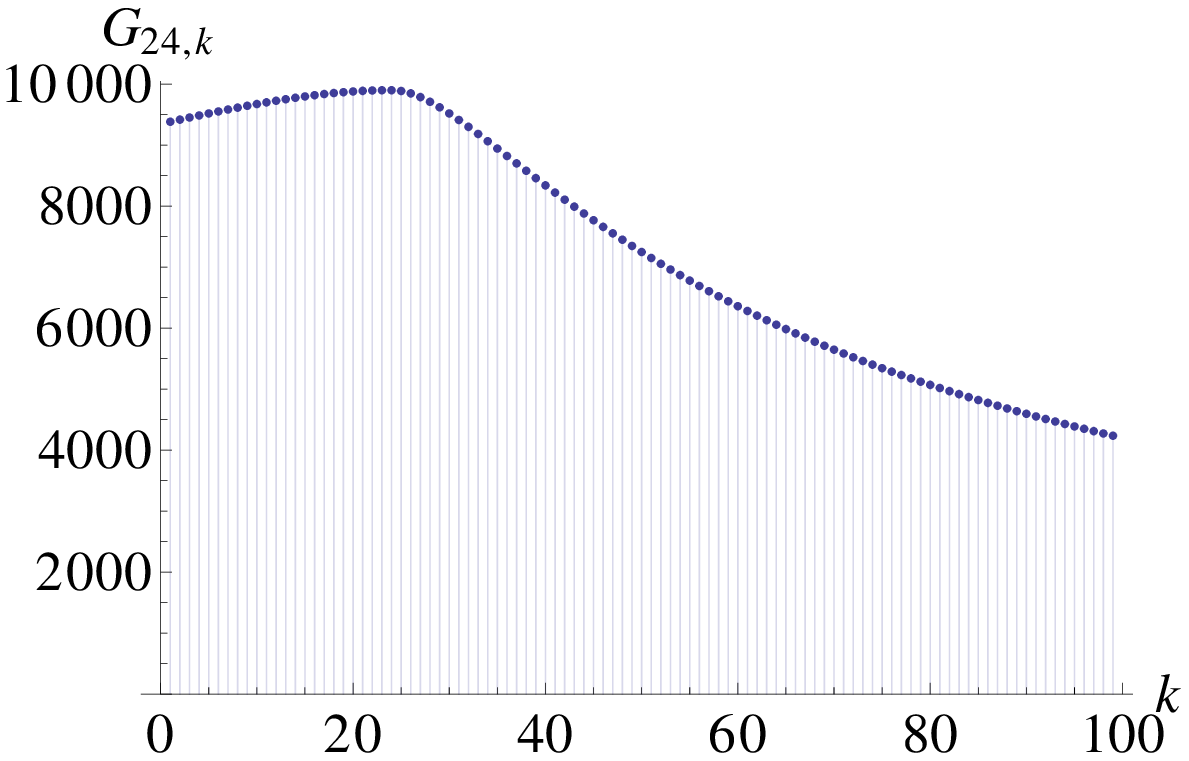}&\includegraphics[width=0.3\textwidth]{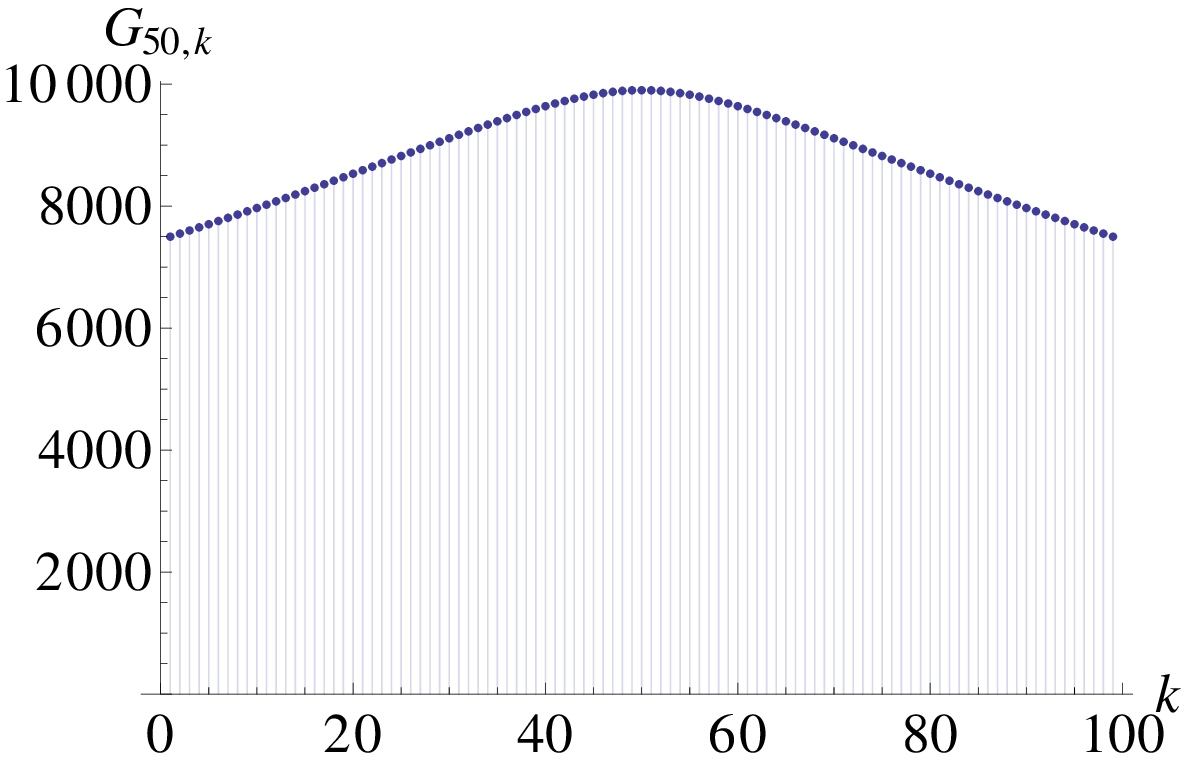}
	\end{tabular}
	\caption{The variance in the number of times a realization starting in $X_1, X_{24}$ and $X_{50}$ is in $X_k$ before absorption as function of $k$ for $N=100$. Notice the scale as compared with Fig. \ref{fig:F_DifferentJ} }
	\label{fig:F2_DifferentJ}
\end{figure}

  Notice that (\ref{eq:Fij1}, \ref{eq:Fij2}) and (\ref{eq:Gij1}, \ref{eq:Gij2}, \ref{eq:Gij3}) can be used to gain new insight into the dynamics inside the transient. $F_{i,k}$ is the mean of the time the process is in the transient configuration $X_k$ when started in the configuration $X_i$ and $G_{i,k}$ is the corresponding variance. Figs. \ref{fig:F_DifferentJ} and  \ref{fig:F2_DifferentJ} show a quite different behavior depending on the initial situation. Starting from $X_i$ close to $X_1$ or $X_{N-1}$ -- the strongly "biased" configurations -- the residence mean times in $X_k$ naturally decrease with the distance from $i$ but become almost independent of $k$ and $N$ for $k$ large whereas the corresponding variance diverges with $N$. Instead, starting from  $X_i$ close to $X_{N/2}$, the quasi-homogeneous configurations- the residence mean times and variance in $X_k$ always diverge.
 \ms

\ms

\ms

The reason for such "strange" behavior is quite clear: as $N$ becomes large, almost all the realizations are trapped during very large times close to their initial configuration, see (\ref{eq:PMat}), and only very few realizations reach the opposite configurations but staying there for large times.
That is, a complete overturn of the opinions is very rare but, when happened, the new situation naturally becomes as stable as the previous.
Therefore we are in a case where there is almost no realization behaving as the mean.
On the other hand, starting from $X_i$ closed to $X_{N/2}$, the "homogeneous" configurations, the mean times in $X_k$ also decrease with the distance from $N/2$, but now the mean times all scale linearly with $N$ and the variances with $N^2$.
Surprisingly these two behaviors, almost static on the border and very unstable "back-and-forth" on the center, compensate perfectly to end up in the same mean residence times and variance (the diagonals of $F$ and $G$) for all the initial configurations.
The same compensation appears when we compare the probabilities for a walk stating in $X_i$ to return in $X_i$, which is independent of $i$ and almost sure for large $N$:

\ms
\begin{equation}\label{eq:pii}
	\lim_{t\rightarrow\infty} p^{(t)}(X_i,X_i) = \frac{F_{ii}-1}{F_{ii}} = \frac{N-1}{N}.
\end{equation}

\ms

Finally the probabilities for a process starting in $X_i$ to end up in $X_0$ or $X_N$ can be computed using also (\ref{eq:Fij1}, \ref{eq:Fij2}). The result is as expected:

\begin{equation}
\lim_{t\rightarrow\infty} p^{(t)}(X_i,X_0) = \frac{N-i}{N}
\label{eq:T2.pXkX0}
\end{equation}
and
\begin{equation}
\lim_{t\rightarrow\infty} p^{(t)}(X_i,X_N) = \frac{i}{N}.
\label{eq:T2.pXkXN}
\end{equation}

\ms

It is reasonable to hypothesize a correlation, if not a causal link, between fast changes in the agent opinion induced by the social process, here stylized in the dynamical rules, and the inconsistency experienced by agents between the micro and the macro level, described in section ~\ref{section-introduction}. This conflict is referred to as practical emergence. It consists of a gradual separation of the individual mental patterns from the reality. The agent is then faced with a representation that is not always perfectly in keeping with the situation  (\cite{Giesen1987}, pag.342).
In the opinion model, a possible rating of this practical emergence inconsistency is the mean time the macro process takes to change of state. Indeed any change of state in this process corresponds to a change for a opinion of an agent . Therefore the faster this rate, the smaller the switching mean time,  and the more likely is the emergence of a practical disruption between picture and reality from the agent's point of view..

\ms

From (\ref{eq:F1T2.p}) and Ref. \cite{Kemeny1976}, Thm. 3.5.6, the mean time $\eta_k$ that the process remains in state $X_k$ once the state is entered (including the entering step) is:

\ms
\begin{eqnarray}\label{eq:meantime}
\eta_k = \frac{N^2}{2k (N-k)}.
\end{eqnarray}

\ms
Therefore, $\eta_k$ is of order $\frac{N}{2}$ for $k$ close to (but smaller than) $N$ and $2$ for $k$ close to $\frac{N}{2}$. Again, for $N$ large the process will be almost stationary in presence of a large majority supporting one of the opinions but extremely unstable when no opinion is clearly predominant. In the latter case practical emergence is plausible. We suggest correlating small values of $\eta_k$ with this phenomenon.

\ms

\begin{figure}[htbp]
	\centering
\includegraphics[width=0.6\textwidth]{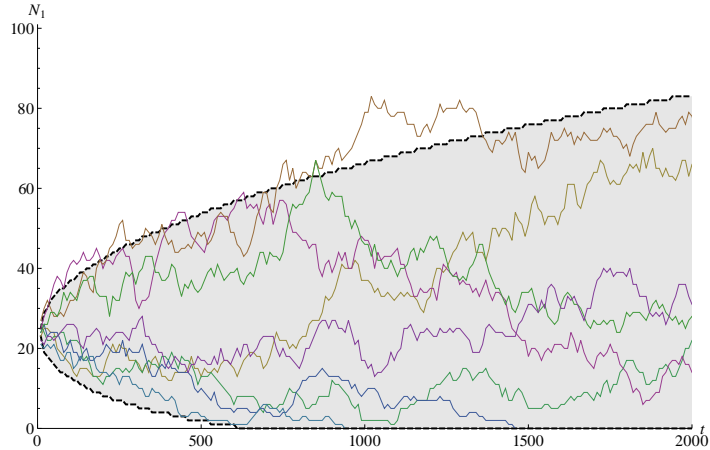}
	\caption{Different realizations of simulations with 24 out of 100 agents in initial state '1' (i.e a process starting in $X_{24}$). Markov chain analysis shows that with probability $0.95$ the process is in the shaded region.}
	\label{fig:TimeEvolution03}
\end{figure}

To conclude this section Fig. \ref{fig:TimeEvolution03} shows different realizations of the agent simulation along with the expected evolution in form of a confidence interval.
The measure of the realizations inside a given confidence interval is an increasing function of time. However, since any individual realization may cross the border of this interval several times before falling in one of the final absorbing states a numerical evaluation of the convergence times may be quite  delicate.

\ms

 \subsection{The Macro Dynamics of a General Opinion Model}~\label{subsection-any-delta}

 \ms

 For an opinion model with $\delta$ different opinions, the opinion of any agent $i$ at time $t$ is a  variable $x_i(t) \in \left\{0,\dots\ , \delta -1\right\}$.
The opinion profile is given by the vector $x(t) = \{x_1(t),\ldots,x_N(t)\}$.
The space of all possible configurations is then $\Sigma = \left\{0, \dots, \delta-1\right\}^N$.
Following the same argument as for $\delta = 2$, we define $N_s (x)$ to be the number of agents in the configuration $x$ with opinion $s$, $s= \left\{0, \dots, \delta - 1\right\}$,  and then $X_{<k_0, k_1, \dots, k_{\delta - 1}>} \subset \Sigma$ as
\begin{equation}
X_{<k_0,  \dots, k_s, \dots, k_{\delta _ 1}>}  = \left\{x \in \Sigma \ : N_0(x) = k_0, \dots, N_s(x) = k_s, \dots, N_{\delta -1} (x) = k_{\delta - 1} \mbox{ and } \ \sum_{s=0}^{\delta - 1}  k_{s} = N\right\}.
\label{eq:X< >}
\end{equation}
Each $X_{<k_0, k_1, \dots, k_{\delta - 1}>}$ contains all the configurations $x$ in which exactly $k_s$ agents hold opinion $s$ for any $s$.
As in the binary case, in this way we obtain a partition of the configuration space $\Sigma$.

\ms

It is clear that also in this case the group of all the permutations of $N$ agents' labeling fulfills conditions (1) and (2) of Proposition(\ref{proposition_lumpability}) and that condition (3) is verified if  the probability distribution $\omega$ is permutation invariant and therefore uniform:  $\omega(i,j) = \frac{1}{N^2}$, for all pair of agents ($i,j$).

\ms

In this case, Eq. (\ref{eq:F1T2.p}) generalizes to:
\begin{equation}
P(X_{<k_0, k_1, \dots, k_{\delta - 1}>}, X_{<k'_0, k'_1, \dots, k'_{\delta - 1} >}) = \frac{k_s k_r}{N^2}
\label{eq:Pp< >}
\end{equation}
if $k'_s = k_s \pm 1$ and $k'_r = k_r \mp 1$ whereas $k'_j = k_j$ for all other $j$, and the probability that no opinion changes, Eq. (\ref{eq:F1T2.q}), becomes
\begin{eqnarray}
P(X_{<k_0, k_1, \dots, k_{\delta - 1} >}, X_{<k_0, k_1, \dots, k_{\delta - 1}>}) = \frac{1}{N^2} \sum_{s=0}^{\delta - 1}  (k_{s})^2\nonumber \\.
\label{eq:Pq< >}
\end{eqnarray} 

\ms

The structure of (\ref{eq:Pp< >}) has an interesting consequence on the dynamics of the system. We see that, if one $j$ for which $k_j =0$, the probability of transition to a state with $k_j = 1$ is $0$. In other words, to change the number of agents sharing opinion $j$, at least one agent with such an opinion is needed. Therefore, the state space is organized as a $ \delta - simplex$  with absorbing faces ordered by inclusion, corresponding to increasing sets of opinions with no supporters.

\ms

Starting in some state with no null $k_j$ the process will finish at certain time in a state where, for the first time, $k_j = 0$ for some $j$ (notice that only one $j$ at each time can fall to zero since the sum of all $k_j$ is constant). From there, the given $k_j$ will stay equal to zero for ever, and (\ref{eq:Pp< >} -- \ref{eq:Pq< >}) tell us that the transition probabilities are now those of a system with $\delta - 1$ opinions. Because the condition $ \sum_{s=0}^{\delta-1}  k_{s} = N$ is to be fulfilled by the remaining opinions, the system will then evolve exactly as if the N agents share $\delta-1$ opinions from the very beginning.  After a certain time a new opinion will lose all its supporters and the system is now equivalent to a full system of $\delta - 2$ opinions, and so on. The system will cascade up to the final absorbing state, with only one opinion shared by all the $N$ agents. We recall that each of such cascade transitions is achieved in finite (random) times.

\ms

By computing the fundamental matrix of the subsystems it would be possible to access the mean and variance of the times the system evolves between two successive extinctions of group opinions. We conjecture the same scaling laws for a system of $\delta$ opinions as the ones already described for $\delta = 2$.

\ms

 \subsection{The Macro Dynamics, Further Reduction}~\label{subsection-reduction}

\ms

Alternatively, we can make use of the symmetries in the structure of (\ref{eq:Pp< >}) and search for lumpable partitions to further reduce the problem.

\ms

This can be done by considering the model from the perspective of a single "party" associated with (say) opinion $0$.
For that "party", it may be important to know how many agents are supportive because they share the same opinion, and how many are not because they support one of the remaining opinions.
Thus, we reduce the model to a quasi--binary variant with the supporter opinion $0$ on one side and and all other opinions ($1 \cup \dots \cup {\delta - 1}$) on the other side, grouping together all the states with $k_0 = r$, $r= 0, \dots, N$.

\ms

The corresponding partition reads:
\begin{equation}
Y^0_r = \bigcup_{\substack{ k_1, \dots, k_{\delta - 1}\\k_0 = r}} X_{\left\langle r, k_1, \dots, k_{\delta - 1}\right\rangle},  r= 0, \dots, N.
\label{eq:F1T3.Yr}
\end{equation}
It is easy to verify that the chain (on the $X$) is indeed lumpable with respect to $Y$ and that
\begin{equation}
P(Y^0_r,Y^0_{r+1})= P(Y^0_r,Y^0_{r-1}) = \frac{r(N - r)}{N^2}.
\label{eq:YLumpProb}
\end{equation}
It thus turns out that the chain formed by the $Y^0_r, r = 0,1,\ldots,N$ is exactly the same as the chain derived for the binary model.
Therefore, the questions regarding the evolution of one opinion in relation to all the others taken together are addressed by the transient analysis performed in Sec.~\ref{subsection-delta2-transient}. That is to say, from this point of view, each "party" may rely on the dynamics of a binary model as a coarse description of the evolution of its own status.

\ms

There is however an important subtlety when doing such an analysis. The asymmetry of the partition one-against-all-others will be encoded in the initial condition. For instance, starting with an equally distributed profile of N agents corresponds to the initial condition $X_{<k, k, \dots, k>}$ in the detailed description but to $Y^0_{N/ \delta}$ in the coarse case. In such a way the asymmetry in the one-against-all-others description is recovered.

\ms

Another important issue concerns the effects of bounded confidence in the model, in other words if a certain number of pairs of opinions do not communicate. From a formal point of view bounded confidence is encoded in the confidence matrix $\alpha$ just by putting  $\alpha (a,b) = 0$ if ($a,b$) is one of such non--communicating opinion pair, denoted to as $a \nleftrightarrow b$.  The consequence of bounded confidence is the emergence of non--consensual absorbing states known as opinion clusters. In the following section, we treat in great detail the simplest case where bounded confidence is possible, namely $\delta= 3$. We postpone the general case to the last section since it is a simple generalization of this example.

 \ms

 \subsection{The Macro Dynamics of a Three Opinions Model and the Emergence of Opinion Clustering}~\label{subsection-delta-three}

 \ms

\begin{figure}[ht]
	\centering
		 \includegraphics[width=.90\textwidth]{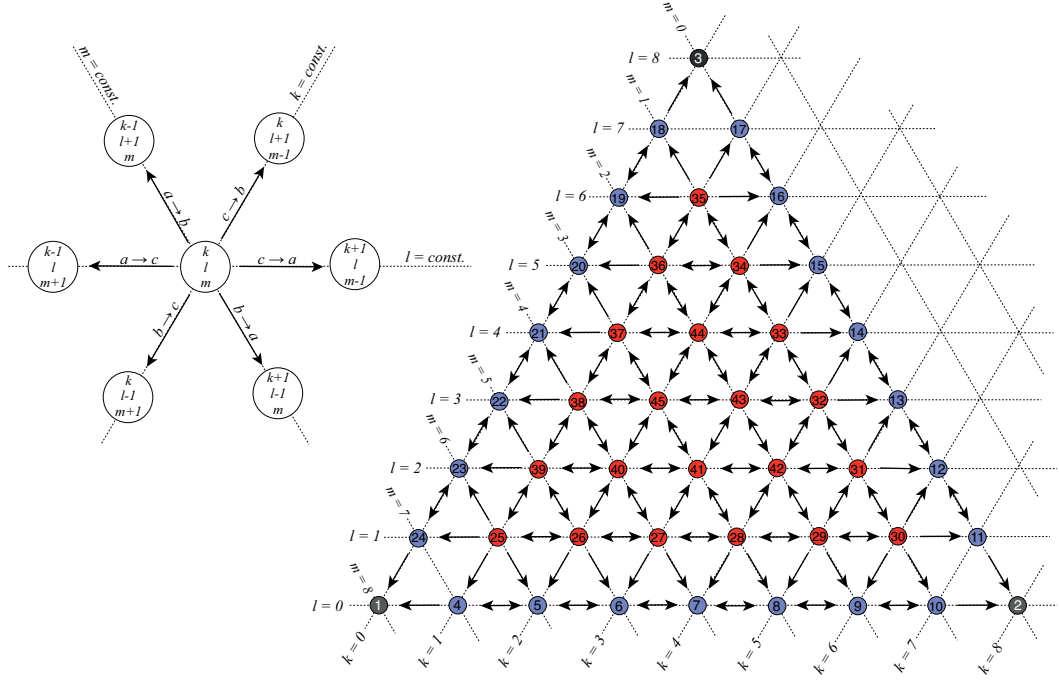}
	\caption{Transition structure (l.h.s) and state topology (r.h.s) of the unbounded confidence model with three opinions \{a, b, c\} , here $N=8$.}
	\label{fig:StateTopologyAndTransitions.T3.N8}
\end{figure}

We are particularly interested in the $\delta = 3$ case because it is the simplest version in which one can meaningfully consider \emph{bounded confidence} effects.
According to the general results of Sec.~\ref{subsection-any-delta}  in case of unbounded confidence, the projection from micro to macro dynamics is  lumpable (under the homogeneous hypothesis on $\omega$ of course).The Markov chain topology obtained by this projection is shown in Fig. \ref{fig:StateTopologyAndTransitions.T3.N8} along with the transition structure.
The probabilities of the transitions are given by Eqs. (\ref{eq:Pp< >}) and (\ref{eq:Pq< >}) which allows us to compute the complete transition matrix $P$.

\begin{figure}[htbp]
	\centering
		\includegraphics[width=.90\textwidth]{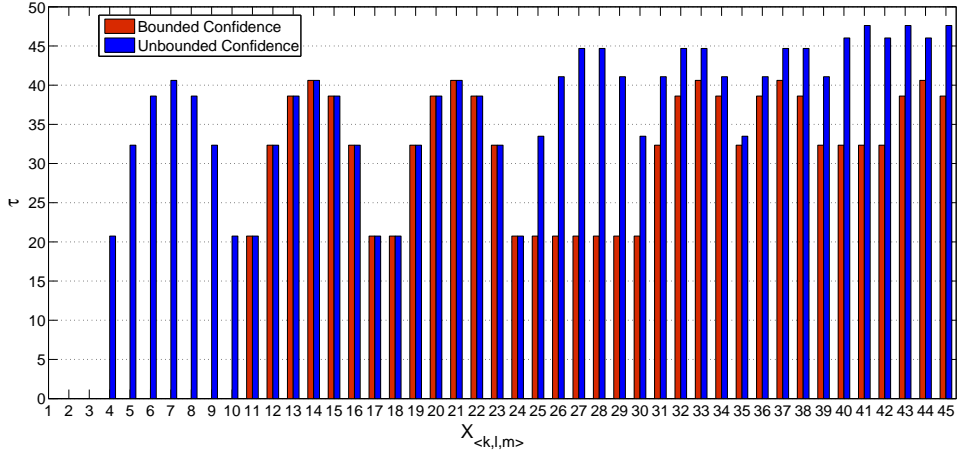}
	\caption{Mean times for the transient nodes to reach an absorbing state. Blue bars: unbounded confidence, red bars: bounded confidence with $a \not\leftrightarrow c$. Labels of the nodes are explained in the text.}
	\label{fig:T3.N8.AbsorbTimes}
\end{figure}

For the construction of $P$, the nodes in the Markov chain are labeled in increasing order from the absorbing to the central nodes, see  Fig. \ref{fig:StateTopologyAndTransitions.T3.N8}: labels 1 to 3 (black) for absorbing consensus states, labels 4 to 24 (blue) for two--opinion states, labels 25 to 39 (red) for three-opinion states with one of the opinion supporters reduced to one element, and labels 40 to 45 (red) for the remainder states.
It is possible to compute the fundamental matrix, at least numerically if $N$ is large, and this makes it possible to compute the significant statistical indicators of the model.
For instance, if $N=8$, the state space of the macro dynamics has 45 states and the mean times for the transient nodes to reach an absorbing state (consensus) range between 21 and 48 time steps, see Fig. \ref{fig:T3.N8.AbsorbTimes}.
Not surprisingly the mean transition times are a function of the distance to the absorbing states as measured on the graph of the state space (Fig.~\ref{fig:StateTopologyAndTransitions.T3.N8}).

\ms

\begin{figure}[htbp]
	\centering
		\includegraphics[width=.90\textwidth]{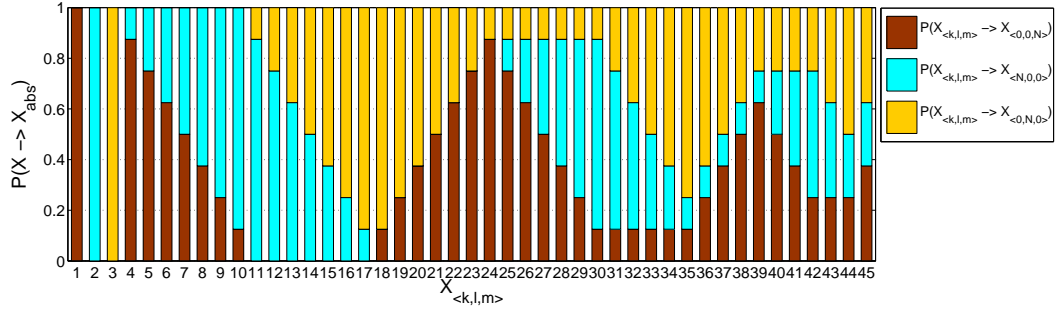}
	\caption{The probabilities of reaching the three absorbing states for all initial nodes $X_{<k,l,m>}$.  Notice that all three final states can be reached only from the inner nodes (numbers 25 to 45).}
	\label{fig:AbsProbs.T3.N8}
\end{figure}

From the fundamental matrix {\bf F}  it is also easy to compute the probabilities of ending up in each of the absorbing (consensus) states starting from any transient node using the matrix $B = {\bf F} R$, where $R$ is defined in (\ref{eq:CanonicalForm}) .
For instance, for $N=8$, the absorbing probabilities for any state are represented in Fig. \ref{fig:AbsProbs.T3.N8}
\ms

\ms

Let's now turn to the question of what happens if agents with a certain opinion do not accept to change their opinion after meeting an agent of another given opinion.
In the opinion dynamics literature, this is referred to as \emph{bounded confidence}.
From the Markov chain perspective the emergence of opinion polarization becomes a simple consequence of the restrictions posed on the interaction process.
As certain transitions are excluded, the state space topology of the Markov chain changes in a way that new absorbing states become present.
The respective states correspond to non--consensus configurations, hence, they represent a population with opinion clustering.

\ms

\begin{figure}[ht]
	\centering
		 \includegraphics[width=.90\textwidth]{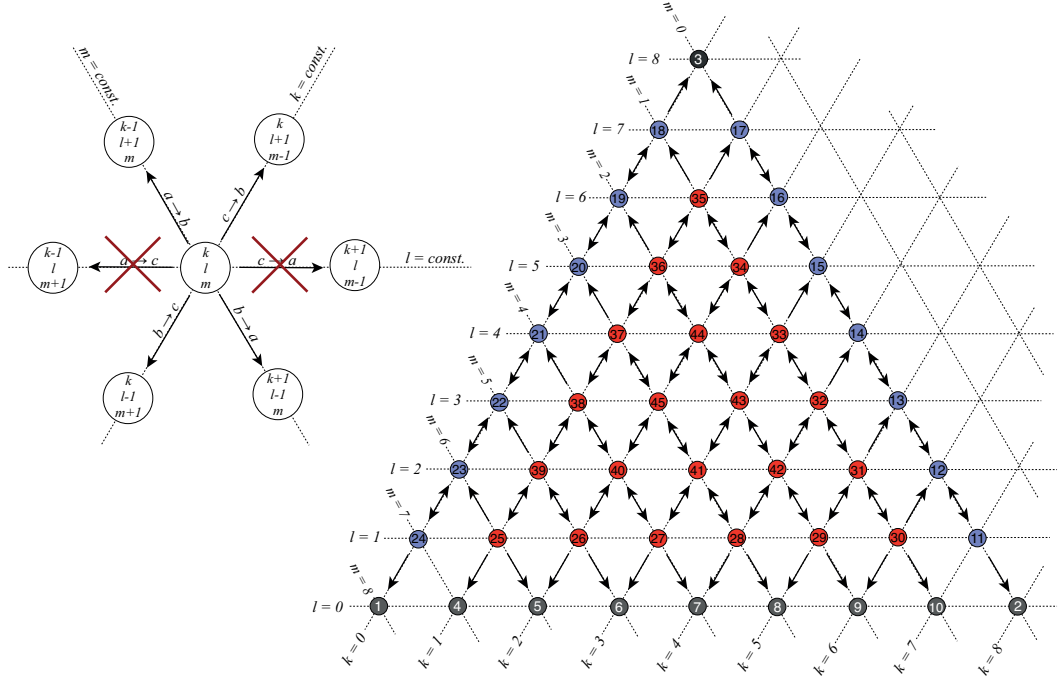}
	\caption{Transition structure and state topology of the bounded confidence model for $N = 8$.}
	\label{fig:StateTopology.BC}
\end{figure}

As an example, let us assume that agents in opinion state 'a' are not willing to communicate with agents in state 'c' and vice versa, that is to say  $\alpha (a,c) =  \alpha (c,a) = 0$.
The corresponding Markov transition matrix $P$ now reads :
\begin{equation}
P(X_{<k,l,m>},X_{<k-1,l,m+1>}) = P(X_{<k,l,m>},X_{<k+1,l,m-1>}) = 0.
\label{eq:F1T3.BC.ProbPairs37}
\end{equation}
and
\begin{equation}
P(X_{<k,l,m>},X_{<k,l,m>}) = \left(\frac{k^2 + l^2 + m^2}{N^2}\right) + 2\left(\frac{k m}{N^2}\right).
\label{eq:F1T3.BC.ProbSelf}
\end{equation}
The remaining entries are, as before, (\ref{eq:Pp< >}) and (\ref{eq:Pq< >}).
The resulting state space topology is shown in Fig.~\ref{fig:StateTopology.BC}, where all horizontal transition paths are removed, since those paths correspond to the  ${a\leftrightarrow c}$ opinion changes.

\ms

For the set of bordering nodes $X_{<k,0,N-k>} : k = 1,\ldots,N-1$ there is no longer any transition that leads away from them, so that all these nodes become absorbing states.
The fact that these additional absorbing states $X_{<k,0,N-k>}$ represent opinion configurations with $k$ agents in state $a$ and $N - k$ agents in state $c$ explains why the introduction of interaction restrictions leads to possible final states with opinion polarization.
It is noteworthy, however, that the opinion clustering would not be observed if only one of the two transitions, ${a\rightarrow c}$ or $c\rightarrow a$, were excluded.
In this case, there would still be a path leading away from the bordering nodes to one of the nodes ($X_{<0,0,N>}$ or $X_{<N,0,0>}$) in the corner of the graph.
Such a set--up corresponds to an asymmetric model where the bordering atoms $X_{<k,0,N-k>} : k = 1,\ldots,N-1$ become again transient, such that the process eventually leads to the final consensus configurations as previously described.
However the final configuration $x = \{\vec{c}\}$ would be much more likely than $x = \{\vec{a}\}$, as a consequence of the asymmetry of such a model variant.

\ms

\begin{figure}[htbp]
	\centering
		\includegraphics[width=.90\textwidth]{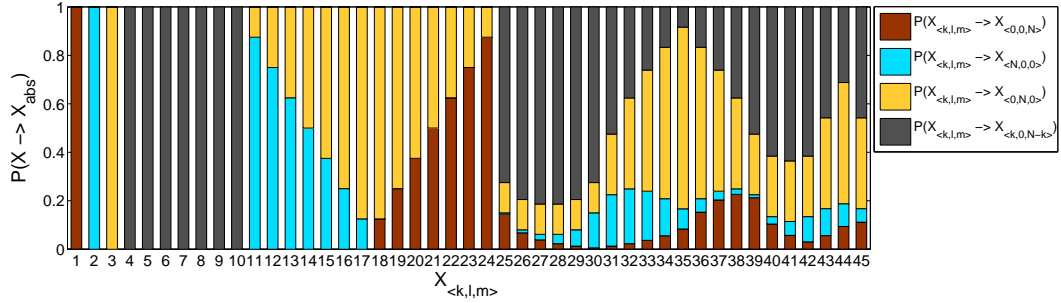}
	\caption{The probabilities for all initial nodes $X_{<k,l,m>}$ converging to opinion clustering or to the three consensus nodes. Notice again that all final states and the non--consensus states in particular can be reached only from the inner nodes (numbers 25 to 45).}
	\label{fig:AbsProbs.T3.N8.BC}
\end{figure}

As for the case of unbounded confidence, the fundamental matrix can be computed here as well allows us to calculate the statistical quantities of the model such as absorbing probabilities and times.
In Fig.\ref{fig:AbsProbs.T3.N8.BC} the probabilities of a realization starting in one of the transient states ending up in each of the absorbing final states are shown for each initial node (computed again by $B = {\bf F} R$).
If the process is in the first 10 nodes at $t=0$, it will remain there forever as all these nodes are absorbing in the bounded confidence case.
Notice that nothing changes for the nodes 11 to 24 with respect to the unbounded case shown in Fig.\ref{fig:AbsProbs.T3.N8}.
For a system in these configurations the communication constraint has no effect on the dynamics.
The six absorbing non--consensus states ( numbers 4 to 10 with only "a" and "c" opinion supporters) are reachable only from the inner nodes, that is only if all opinions are present initially.
It becomes clear that for some of these configurations, the probability of converging to consensus becomes very small (e.g. nodes 25 to 30).

Finally, we can compare the mean time before a realization starting in a transient state remains in the transient before absorption for the bounded and the unbounded case.
This statistical indicator is represented in Fig. \ref{fig:T3.N8.AbsorbTimes}.
Notice that the times for the states 1 to 3 (unbounded) and 1 to 10 (bounded) are zero as in this case the process is absorbed from the very beginning.
Again, the non--absorbing two--opinion states (11 to 24) are not affected.

\ms

As in the general case of any $\delta$  we can search here for lumpable partitions to further reduce the problem taking the point of view of each "party" associated with opinions "a", "b" or "c". For the case of unbounded confidence we have shown in Sec.~\ref{subsection-reduction} that the dynamics from any of these points of view reduces to the $\delta=2$ case.
The status of the bounded confidence model is different. From the perspective of opinion "b" the partition in "supporters" and "opponents" is lumpable, therefore, the system evolves as a binary chain. This is not the case from the perspectives of opinions "a" or "c". For instance, from the point of view of opinion "a", the corresponding partition reads:

\begin{equation}
Y^a_r = \bigcup_{\substack{ l+m = N-r}} X_{\left\langle r, l,m \right\rangle}, \ \ r=0, 1,\dots, N.
\label{eq:F1T3.Ya}
\end{equation}
and
\begin{equation}
P(X_{\left\langle r, l,m\right\rangle}, Y^a_{r+1})= \frac{r l}{N^2}.
\label{eq:YaNoLumpProb}
\end{equation}

\ms
It turns out that the chain formed by the $Y^a_r, r = 0,1,\ldots,N$ is not a Markov chain since the r.h.s. of (\ref{eq:YaNoLumpProb}) depends on $l$ and not only on $r$ \cite{Kemeny1976}.

\ms

We see that the introduction of bounded confidence in this model leads to memory effects due to the fact that an agent switching from opinion "a" to opinion "c" necessarily goes through a visit to opinion "b" for at least one time step, therefore, the probability of this transfer will depend on the number of supporters of opinion "b" at that time.

 \ms

\section{Simple Generalizations}~\label{section-generalizations}

 \ms
 
We first mention an easy generalization of the existence of absorbing states for the case of bounded confidence in a model with any number $\delta$ of different opinions. In order to get non consensual absorbing states it is necessary and sufficient that a subset of opinions is mutually incommunicable. In this case all the states belonging to the simplex generated by the mutually incommunicable opinions become absorbing. It is worthwhile noticing that absorbing states may appear in different clusters of simplexes provided that the corresponding opinions are related by chains of communicating links. An example of this type appears for $\delta = 3$ if ($a \nleftrightarrow b$) and ($a \nleftrightarrow c$) but ($b \leftrightarrow c$) where the absorbing states are either the simplex with only $a$ and $b$ or with only $a$ and $c$ opinions.

\ms

Another interesting issue concerns agent models with vectorial (or equivalently matrix or table) individual state (attribute) space. Suppose that at each time step each $i$ is characterized by a list of $n$ attributes, where the first attribute may take $n_1$ possible values, $\cdots$, and the $n^{th}$ attribute $n_n$ possible values. The corresponding ABM can then be easily built as in Sec.~\ref{subsection-any-delta} by taking $\delta = n_1 \times n_2  \times \cdots  \times n_n$. And, as long as one is interested in following the macrodynamics of the agents who all have $\delta$ attributes that are identical,, the reduction proposed in Sec.~\ref{subsection-reduction} also applies. Therefore, absorbing non consensual states will appear in exactly the same way as described above as a consequence of bounded confidence.

\ms

For this vectorial opinion model there is, however, an unexpected subtlety when we are interested in the macrodynamics of the agents ranked by only one of their attributes, for instance if the agents are separated in $n_1$ different groups according to the number of agents  sharing their first attribute. Then the partition is no longer lumpable, and therefore the evolution of the correspondent random variables (for instance, the number of elements of each group) is not a Markov chain. Again, in this case, new memory effects may appear from this choice of aggregation to build the macrodynamics. The proof can be done as in (\ref{eq:F1T3.Ya}) and (\ref{eq:YaNoLumpProb}).

\ms

We shall finish this section by noticing a trick that makes it possible to extend the previous analysis to systems where the agents are defined on a lattice or, more generally, on a graph. They may coexist in some nodes of the graph or, sometimes they may even move on it. Then, in some cases, it is possible to reduce the model to the framework of our work by defining the agents as the nodes in the graph and including the presence or lack of an actor in the prescription of the attributes defining the elements of $\bf{S}$ as well as the topology of the graph accordingly in$\alpha$.

\section{Final Remarks and Prospectives.}~\label{section-conclusion}
\ms

An important mark of ABMs is their ability to include arbitrary levels of heterogeneity and stochasticity (or uncertainty) into the description of a system of interacting agents.
While computer simulations are often suited for making important dynamical trends of these models visible, a rigorous characterization of the different dynamical phases is difficult.

\ms

In this work we analyze the dynamics of an ABM from a Markovian perspective and derive explicit statements about the possibility of linking a microscopic agent model to the dynamical processes of macroscopic observables that are useful for a precise understanding of the model dynamics. In this way the dynamics of collective variables may be studied, and a description of macro dynamics as emergent properties of micro dynamics, in particularly during transient times, is possible. In our context, the random map representation (RMR) of a Markov process helps to understand the role devoted to the collection of (deterministic) dynamical rules used in the model from one side and of the probability distribution $\omega$ governing the sequential choice of the dynamical rule used to update the system at each time step from the other side. The importance of this probability distribution, often neglected, is to encode the design of the social structure of the exchange actions at the time of the analysis.  Not only, then, are features of this probability distribution concerned with the social context the model aims to describe, but also they are crucial in predicting the properties of the macro--dynamics. If we decide to remain at a Markovian level, then the partition, or equivalently the collective variables, to be used to build the model may be compatible with the symmetry of the probability distribution $\omega$. In a sense the partition of the configuration space defining the macro--level of the description has to be refined in order to account for an increased level of heterogeneity or a falloff in the symmetry of the probability distribution. It is, however, clear that, in absence of any symmetry, there is no other choice for this partition than to stay at the micro--level and in this sense, no Markovian description of a macro--level, is possible in this case.

\ms

At this point two possible approaches could be explored in the future: first to identify in specific models other natural symmetries inducing a Markovian description of their macro counterpart and to study the evolution of the corresponding collective behavior; and second, to tackle the problem of describing the dynamics of the macro variables in cases where they are not Markovian. For this purpose the cases where $\omega$ is not  homogeneous (or symmetric) but has special structural properties seem to be a crucial issue for further research. This point must have an important impact in the understanding of descriptive emergence since it is a (the, in ABM context) source of long memory effects in the dynamics of the formation of collective social patterns.

\ms

 Another important prospect concerns the measure of practical emergence or discrepancy, the gap between the macro-structural properties of a social system and internalized rules or intentions of the individual actors. The measure of this gap should lead to more elaborate gauges whose dynamics themselves call for new specific investigation.

 \ms
 
 As pointed out  in section \ref{section-introduction}, under certain circumstances the macro process may undergo dynamical changes in its own structural rules. This fact is referred to as explanatory emergence. It can be understood either as a consequence of some external (to the model) inputs or on the basis of deep accelerations of the micro dynamics that in turn bring about the processes of change at the macro level. In both cases this question opens up to new theoretical as well as very interesting practical developments.

\ms

 In an ABM as presented in section ~\ref{section-microdynamics}, the macro-structural patterns of social relations are encoded in the probability distribution $\omega$, or its generalization when the model allows a simultaneous updating depending on more than two agents. Therefore a dynamics of this structure may be incorporated in the model as a dynamics of the probability distribution, allowing it to be time dependent: $\omega_t$. This can be done at different levels. The simplest one occurs when the structural dynamics is assumed to be autonomous, in the form of a master-slave system where the dynamics of the probability distribution is defined independently of the evolution of the process. It is natural to suppose that this dynamics is slow with respect to a fast dynamics of the agent changes. A more sophisticated modeling consists of coupling the individual agent dynamics with the dynamics of the distribution. This corresponds to assuming a feedback of the agent dynamics on the evolution of the structural rules fixing the macro dynamics. In both cases a rigorous treatment of the problem is  sufficiently compelling to deserve further research.

\ms

The formalization of the relations between the micro and the macro levels in the description of the
dynamics of Agent Based Models as well as their mathematical characterization is a step towards
a mathematical theory of emergence in complex adaptive systems. In this work we showed how a
Markov chain approach, in particular the use of the notion of lumpability, provides useful instruments
for the analysis of the link from a microscopic ABM to macroscopic observables. Further research is
needed to deepen our understanding of this link in a more general setting.
\ms

\section{Acknowledgments}

We are grateful to Ann Henshall for a careful reading of the manuscript and to Dima Volchenkov for helpful discussions.
Financial support of the German Federal Ministry of Education and Research (BMBF) through the project \emph{Linguistic Networks} ({\tt http://project.linguistic-networks.net}) and the Portuguese Foundation for Science and Technology (FCT) through the project PDCT/EGE/60193/2004 is also gratefully acknowledged.


\end{document}